\documentclass[fleqn,usenatbib]{mnras}

\usepackage[T1]{fontenc}

\DeclareRobustCommand{\VAN}[3]{#2}
\let\VANthebibliography\thebibliography
\def\thebibliography{\DeclareRobustCommand{\VAN}[3]{##3}\VANthebibliography}

\usepackage{graphicx}	
\usepackage{amsmath}	
\usepackage{amssymb}	
\usepackage{multirow}
\usepackage{subfig}

\usepackage{newtxtext,newtxmath}

\title[RealSim-IFS]{Realistic synthetic integral field spectroscopy with RealSim-IFS}

\author[Bottrell et al.]{
Connor Bottrell$^{1}$\thanks{E-mail: connor.bottrell@ipmu.jp} \& Maan H. Hani$^{2}$ \\
$^{1}$Kavli Institute for the Physics and Mathematics of the Universe (WPI), UTIAS, University of Tokyo, Kashiwa, Chiba 277-8583, Japan\\
$^{2}$Department of Physics and Astronomy, McMaster University, Hamilton, Ontario L8S 4M1, Canada\\
}

\date{Accepted XXX. Received YYY; in original form ZZZ}

\pubyear{2015}

\begin{document}
\def \nuprocess{$\nu$-process}
\def \nodata{. . .}
\def \degree{$^{\circ}$}
\def \Msolar{M$_{\odot}$}
\def \alphafe{[$\alpha$/Fe]}
\def \na{New Astronomy}
\def \HI{H\ion{I}}
\def \sion{\ion{II}}
\def \vninety{v$_{90}$}
\def \Lbol{L$_{\rm bol}$}
\def \Mstar{M$_{\star}$}
\def \logMstar{\log($M$_{\star}/$M$_{\odot})}
\def \logRimp{log(R$_{\rm imp}$/kpc)}
\def \logLbol{log(L$_{\rm AGN}$/erg s$^{-1}$)}
\def \logsSFR{log(sSFR / yr$^{-1}$)}
\def \kms{km s$^{-1}$}
\def \zabs{z$_{\rm abs}$}
\def \zem{z$_{\rm em}$}
\def \Rimp{$\rho_{\rm imp}$}
\def \Rvir{$\rho_{\rm vir}$}
\def \deltaEW{${\rm \Delta log(EW/m\AA)}$}
\def \RadRat{$f_{\rm AGN}/f_{\rm HM01}$}
\def \mnfe{[Mn/Fe]$_{\rm DC}$}
\def \Mgeqw{W$_{0}^{2796}$}
\def \Feeqw{W$_{0}^{2600}$}
\def \fracMgFe{ ${\rm{W}_{0}^{2796}}$/${\rm{W}_{0}^{2600}}$}
\def \omegaDLA{$\Omega_{\rm H \textsc{i}}$}
\def \ndla{30}
\def \npdla{46}
\def \nlpdla{41}
\def \nxpdla{5}
\def \nmdla{27}
\def \nlmdla{21}
\def \nxmdla{6}
\def \CosmoZ{$\langle Z/Z_{\odot} \rangle$}
\def \fNX{$f(N,X)$}
\newcommand{\gimtwod}{\textsc{gim2d}}
\newcommand{\sersic}{s\'{e}rsic} 
\newcommand{\Sersic}{S\'{e}rsic}
\newcommand{\sextractor}{\textsc{SExtractor}}
\def \rifs{\textsc{RealSim-IFS}}
\def \realsim{\textsc{RealSim}}
\def \tpost{T_{\mathrm{PM}}}

\label{firstpage}
\pagerange{\pageref{firstpage}--\pageref{lastpage}}
\maketitle

\begin{abstract}
The most direct way to confront observed galaxies with those formed in numerical simulations is to forward-model simulated galaxies into synthetic observations. Provided that synthetic galaxy observations include similar constraints and limitations as real observations, they can be used to (1) carry out even-handed comparisons of observation and theory and (2) map the \emph{observable} characteristics of simulated galaxies to their \emph{a priori} known origins. In particular, integral field spectroscopy (IFS) expands the scope of such comparisons and mappings to an exceptionally broad set of physical properties. We therefore present \rifs{}: a tool for forward-modelling galaxies from hydrodynamical simulations into synthetic IFS observations. The core components of \rifs{} model the detailed spatial sampling mechanics of any fibre-bundle, image slicer, or lenslet array IFU and corresponding observing strategy, real or imagined, and support the corresponding propagation of noise adopted by the user. The code is highly generalized and can produce cubes in any light- or mass-weighted quantity (e.g. specific intensity, gas/stellar line-of-sight velocity, stellar age/metallicity, etc.). We show that \rifs{} exactly reproduces the spatial reconstruction of specific intensity and variance cubes produced by the MaNGA survey Data Reduction Pipeline using the calibrated fibre spectra as input. We then apply \rifs{} by producing a public synthetic MaNGA stellar kinematic survey of 893 galaxies with $\logMstar>10$ from the TNG50 cosmological hydrodynamical simulation. 
\end{abstract}



\section{Introduction}

The current generation of multi-object integral field spectroscopy (IFS) programs have made it possible to connect the internal properties of large, heterogeneous samples of galaxies to their global characteristics and environments (e.g. the Calar Alto Legacy Integral Field Area (CALIFA, \citealt{2012A&A...538A...8S}): $600$ galaxies, the Sydney-AAO Multi-object IFS survey (SAMI, \citealt{2012MNRAS.421..872C}): $3,600$ galaxies, and the Mapping Nearby Galaxies at Apache Point Observatory survey (MaNGA, \citealt{2015ApJ...798....7B}): $10,000$ galaxies). The continued growth in the data quality and scope of IFS surveys shows no sign of slowing, with new instruments staged to begin observing in the near future that will double the number of galaxies with IFS data in the local Universe (e.g. Hector, \citealt{2020SPIE11447E..15B}: $15,000$ galaxies). 

These observational IFS programmes and their predecessors have benefited from a high degree of synergy with galaxy formation models born out in hydrodynamical simulations. Explicit tracking of the assembly and evolution of baryonic content makes hydrodynamical simulations uniquely equipped to diagnose the origins of observed relations and properties of galaxies -- particularly provided cosmological context of their local and large-scale environments. For example, simulations have been used to gain insight into kinematic transformations of galaxies and the origins of fast and slow rotators (e.g. \citealt{2014MNRAS.444.3357N,2017MNRAS.468.3883P,2017ApJ...837...68C,2018MNRAS.473.4956L,2022MNRAS.509.4372L,2018MNRAS.480.4636S}), kinematic misalignments of gas and stars (e.g. \citealt{2015MNRAS.451.3269V,2019ApJ...878..143S,2020MNRAS.495.4542D,2021MNRAS.500.3870K}), the sizes, shapes, and pattern speeds of stellar bars (e.g. \citealt{2021MNRAS.508..926R,2021A&A...655A..97L,2022arXiv220108406F}), resolved star formation profiles and quenching timescales (e.g. \citealt{2017ApJ...849L...2O,2019ApJ...874L..17S,2020MNRAS.494.6053A,2021MNRAS.508..219N,2022MNRAS.511.6126W}), among many others.

These connections between observation and theory hinge on the commonality of features exhibited by observed and simulated galaxies -- which first requires that observational and simulation data are placed on even ground in terms of data quality. The \emph{inverse} approach of extracting the intrinsic properties (e.g. positions and velocities of individual stars) of observed galaxies is generally complicated by limitations arising from observing conditions and instrumental specifications (e.g. aperture characteristics, point-spread function, spectral resolution, line-spread function, etc.). On the other hand, \emph{forward}-modelling of simulations to synthetic observations which incorporate these nuisances is more practical and still yields the desired outcome -- observed and simulated data which can be (1) compared even-handedly and (2) analyzed using the same observational methods. This approach has been used extensively to examine the degree of agreement between morphologies of galaxies found in observations and cosmological simulations (e.g. \citealt{2015MNRAS.454.1886S,2017MNRAS.467.1033B,2017MNRAS.467.2879B,2018ApJ...853..194D,2019MNRAS.483.4140R,2019MNRAS.489.1859H,2022MNRAS.511.2544D}) using models sensitive to even small differences in galaxy substructure \citep{2021MNRAS.501.4359Z}. 

In particular, \cite{2019MNRAS.490.5390B} and \cite{2021arXiv211100961C} explicitly show that accurate statistical representation of observational nuisance factors in the forward-modelling process is essential for the transferability of deep learning models from synthetic to real data, and vice-versa. In short, deep models must be trained on data which are statistically generalizable to the data set to which the models are applied. This hard requirement extends from images to summary statistics of galaxy morphologies (\Sersic{} indices, disk fractions, asymmetries) -- for which measurements can be affected strongly by data quality (e.g. signal-to-noise) and choices in pre-processing, models, and optimization routines (e.g. \citealt{2011ApJS..196...11S,2013MNRAS.430..330H,2014ApJS..210....3M,2014MNRAS.443..874B,2015MNRAS.446.3943M,2019MNRAS.486..390B,2021MNRAS.507..886T}). Therefore, the second key advantage of: (a) accurately incorporating observational effects (\emph{realism}) into synthetic observables and (b) using the same tools to analyze the observations and simulations, is that any divergences between recovered properties will be due to their intrinsic differences and not caused by e.g. an omission of some observational constraint or effect in the synthetic data.

Most detailed forward modelling approaches have thus-far \emph{mostly} focused on imaging. However, the required compatibility between synthetic and observed images must extend to IFS, which is simply an extension of imaging for some number of wavelength or frequency channels. Consequently, the compatibility must hold for all IFS-derived quantities (e.g. velocity and velocity dispersion maps, stellar age/metallicity, gas profiles/metallicities, line strengths). Existing tools for producing realistic IFS data cubes from hydrodynamical simulations include the \textsc{SimSpin}\footnote{\url{https://github.com/kateharborne/SimSpin}}\citep{2020PASA...37...16H} and \textsc{MARTINI}\footnote{\url{https://kyleaoman.github.io/martini}} \citep{2019ascl.soft11005O}. \textsc{MARTINI} is a Python software that produces synthetic spatially-resolved HI line observations (cubes) from mock radio array interferometry and includes modules for the beam, noise, and various models for HI line spectra. \textsc{SimSpin} is an R-package that produces a synthetic stellar kinematic velocity cube with dimensions $(x,y,v_{\mathrm{los}})$, where $v_{\mathrm{los}}$ is velocity along a given line-of-sight. The code includes components for providing the effective\footnote{\emph{Effective} seeing refers to the seeing contributions from atmosphere and all other pre- focal plane components and effects, including the sizes of fibre apertures, the IFU design, and observing pattern.} seeing convolution, incorporation of a line-spread function (LSF) to each particle, and the shape of the observational footprint. 

In addition, tools have been developed to emulate the data characteristics of specific surveys. To examine the ability to recover stellar masses, ages gradients, and the evolution thereof in MaNGA IFU observations, \citealt{2019MNRAS.483.4525I} produced and analyzed synthetic spectral datacubes of two simulated Milky Way-like galaxies for which they carried out a rigorous emulation of a MaNGA IFU system response. The \textsc{SEDmorph-YZCube} code\footnote{\url{https://github.com/SEDMORPH/YZCube}} was used to generate synthetic stellar spectra cubes and incorporate components of the MaNGA atmospheric and instrumental responses for a suite of merger simulations \citep{2020MNRAS.498.1259Z}. Therein, \citealt{2020MNRAS.498.1259Z} showed that even small differences in the algorithms used to emulate the MaNGA instrument system effects yield large differences in the output data characteristics. Most recently, \citealt{2022arXiv220311575N} produced synthetic cubes in which simulation particles were assigned spectra built from real calibrated MaNGA stellar spectra (MaStar stellar library; \citealt{2019ApJ...883..175Y,2020MNRAS.496.2962M,2022ApJS..259...35A}). Given that these synthetic spectra were assembled from real spectra taken with fibres from the MaNGA spectrograph, \citealt{2022arXiv220311575N} were able to incorporate noise into the synthetic spectra using the signal-to-noise characteristics of real MaNGA galaxy observations. These tools are part of a crucial push towards synthetic IFS data with precise matching to real IFU data characteristics.

In this work we present a generalized tool, \rifs{}\footnote{\url{https://github.com/cbottrell/realsim_ifs}}, which expands upon existing tools by enabling near-exact modelling of the instrumental sampling mechanics for any IFS instrument, real or imagined -- including fibre bundles, image slicers, and lenslet arrays. Indeed, we will show that \rifs{} is sufficiently precise in its emulation of fibre-based IFS data reduction pipelines that it can be used to exactly reproduce real flux and variance cubes from the corresponding raw spectra. \rifs{} is highly generalizable and can be used to incorporate an instrument's sampling mechanics into cubes for which the non-spatial dimension is wavelength, gas/stellar velocity (LOSVD cube), stellar age or metallicity, star formation rate, or any other mass- or light-weighted quantity of simulation particles and cells. Most importantly, \rifs{} can be used to generate survey-realistic IFS data from cubes generated by dusty radiative transfer codes including \textsc{sunrise}\footnote{\url{https://bitbucket.org/lutorm/sunrise}} \citep{2006MNRAS.372....2J,2010MNRAS.403...17J} and SKIRT\footnote{\url{https://skirt.ugent.be}} \citep{2015A&C.....9...20C,2020A&C....3100381C} -- which can optionally incorporate kinematics in ray-tracing of stellar light through gas and dust media (e.g. see \citealt{2021ApJ...912...45N}). To emulate the characteristics of data produced by existing fibre-based instruments, \rifs{} makes rigorous use of documentation for the technical specifications, data reduction, and data analysis pipelines of current multi-object IFS surveys (e.g. CALIFA: \citealt{2011A&A...534A...8M}, SAMI: \citealt{2015MNRAS.447.2857B,2015MNRAS.446.1551S}, MaNGA: \citealt{2016AJ....152...83L,2019AJ....158..231W,2021ApJ...915...35L}).

To test and demonstrate the application of \rifs{} to a hydrodynamical simulation, we create a synthetic MaNGA stellar kinematic survey of galaxies from TNG50 cosmological magnetohydrodynamical simulations \citep{2018MNRAS.473.4077P,2018MNRAS.475..648P,2018MNRAS.475..624N}. We make the corresponding fibre-measured LOSVDs and reconstructed cubes publicly available. These data products incorporate the effects of the atmospheric seeing, integral field unit (IFU) fibre characteristics and set-up (designs), dithered exposure strategy, line-spread function, and spatial reconstruction of fibre measurements -- all handled by \rifs{}. To offer high generalizability in the application of these data, sky noise and source poisson noise are omitted. Noise can be incorporated into the stellar velocity data at the user's discretion (e.g. corresponding to desired S/N). However, for any desired treatment of noise, \rifs{} provides built-in support for the accurate propagation of input noise to individual fibre variance spectra/profiles and output variance cubes. Additional testing is carried out with real fibre data from the MaNGA survey to examine the specific continuity that can be achieved between the \rifs{} pipeline and the MaNGA data reduction pipeline \citep{2016AJ....152...83L}. 

This paper is laid out as follows. Section \ref{sec:realifs_data} describes the TNG50 simulation, data products, and the real MaNGA observations that are used in testing and producing our synthetic MaNGA stellar kinematic survey. In particular, the characteristics of the data products outlined in Section \ref{sec:realifs_data} provide the basis for shaping the methods -- which are summarized in Section \ref{sec:realifs_methods}. Section \ref{sec:realifs_results} presents the results of applying \rifs{} to the simulations and our tests with real MaNGA data. Finally, we describe the set-up and data products of a mock MaNGA stellar kinematic survey for galaxies from the TNG50 simulation. The calculations in this chapter adopt a Planck-based cosmology with $H_0 = 67.74$ km s$^{-1}$ Mpc$^{-1}$, $\Omega_{\mathrm{m}} = 0.308$, and $\Omega_{\Lambda}=0.692$ \citep{2016A&A...594A..13P}.

\section{Data}
\label{sec:realifs_data}

We carry out tests and demonstrate the application of \rifs{} to simulated galaxies using the TNG50-1 run from the IllustrisTNG suite of cosmological magneto-hydrodynamical simulations \citep{2018MNRAS.479.4056W,2018MNRAS.473.4077P,2018MNRAS.475..648P,2018MNRAS.475..676S,2018MNRAS.477.1206N,2018MNRAS.480.5113M,2018MNRAS.475..624N,2019ComAC...6....2N,2019MNRAS.490.3234N}. The simulation suite, physical model, and sample selection are laid out in Section \ref{sec:realifs_tng}. Our demonstrations use stellar line-of-sight velocity distribution (LOSVD) cubes for the TNG50 sample. Construction of these cubes and numerical compatibility with cubes measuring other mass- and light-weighted quantities (e.g. spectra) are outlined in Section \ref{sec:realifs_losvd}. Finally, certain tasks within \rifs{} can be used on real IFS survey data (for example, the spatial reconstruction of fibre intensities onto a Cartesian grid) and validated against the final data products from these surveys. These validation tests are carried out using data from the MaNGA survey \citep{2015ApJ...798....7B,2015AJ....149...77D,2015AJ....150...19L,2016AJ....152...83L,2019AJ....158..231W}, summarized in Section \ref{sec:realifs_manga}.

\subsection{IllustrisTNG}
\label{sec:realifs_tng}

Our demonstrations with \rifs{} use the publicly available data from the IllustrisTNG suite of simulations\footnote{\url{https://www.tng-project.org/data/}} \citep{2018MNRAS.479.4056W,2018MNRAS.475..648P,2018MNRAS.473.4077P,2018MNRAS.475..676S,2018MNRAS.477.1206N,2018MNRAS.480.5113M,2018MNRAS.475..624N,2019MNRAS.490.3234N,2019ComAC...6....2N}. IllustrisTNG comprises a suite of large-volume cosmological magneto-hydrodynamical simulations of galaxy formation. The simulations solve for the coupled evolution of dark matter, gas, stars, supermassive black holes, and magnetic fields using the AREPO moving-mesh hydrodynamic code \citep{2010MNRAS.401..791S} and subgrid models including treatments for star formation and stellar evolution, black hole growth, radiative cooling, stellar and black hole feedback, and treatment of magnetic fields. The fiducial TNG simulations are run in three cubic volumes, $V_{\mathrm{box}}=(51.7^3,\;110.7^3,\;302.6^3)$ Mpc$^3$, with descending levels of resolution. The Plummer-equivalent gravitational softening lengths for stellar particles, $\epsilon_{\star}$, and mean star-forming gas cell sizes, $\overline{r}_{\mathrm{cell,SF}}$, in these fiducial simulations are $\epsilon_{\star} = (0.29,\;0.74,\;1.48)$ kpc and $\overline{r}_{\mathrm{cell,SF}}=(0.14,\;0.36,\;0.72)$ kpc, respectively (see Table 1 in \citealt{2019ComAC...6....2N}). All hydrodynamical simulations in the TNG suite use the same physical model described by \cite{2018MNRAS.479.4056W} and \cite{2018MNRAS.475..648P} -- which builds on the original Illustris model \citep{2013MNRAS.436.3031V,2014MNRAS.438.1985T,2014Natur.509..177V,2014MNRAS.445..175G}. We use the highest-resolution run for the $50.7^3$ Mpc$^3$ volume (TNG50-1) for testing with \rifs{}. Given that our main demonstration of \rifs{} is to produce MaNGA-realistic synthetic stellar kinematic data, TNG50-1 is a good candidate because it offers (1) a diverse and realistic galaxy sample (e.g., \citealt{2021MNRAS.501.4359Z}) and (2) baryonic spatial resolution that oversamples the physical resolution of galaxy IFS surveys in the local Universe ($\overline{r}_{\mathrm{cell,SF}}=0.14$ kpc and $\epsilon_{\star}=0.29$ kpc compared to $1-2$ kpc for, e.g., SAMI and MaNGA targets; \citealt{2015MNRAS.447.2857B,2017AJ....154...86W}).

\begin{figure*}
\centering

	\includegraphics[width=0.78\linewidth]{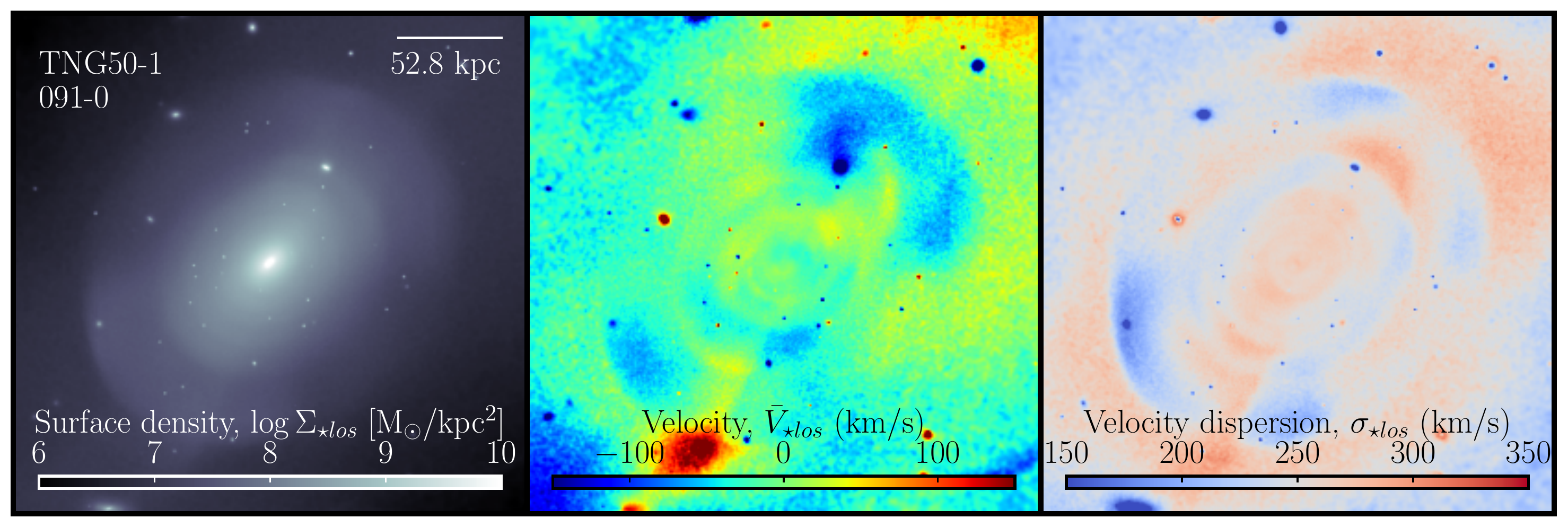}\\
	\vspace{-7pt}

	\includegraphics[width=0.78\linewidth]{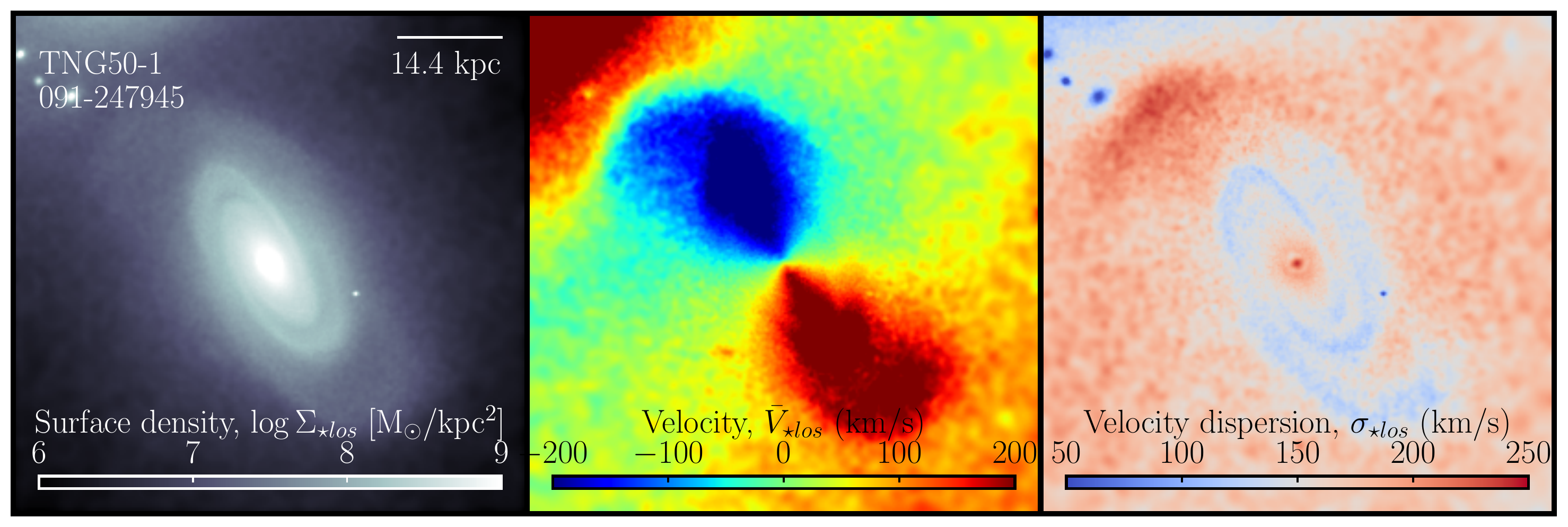}\\
	\vspace{-7pt}

	\includegraphics[width=0.78\linewidth]{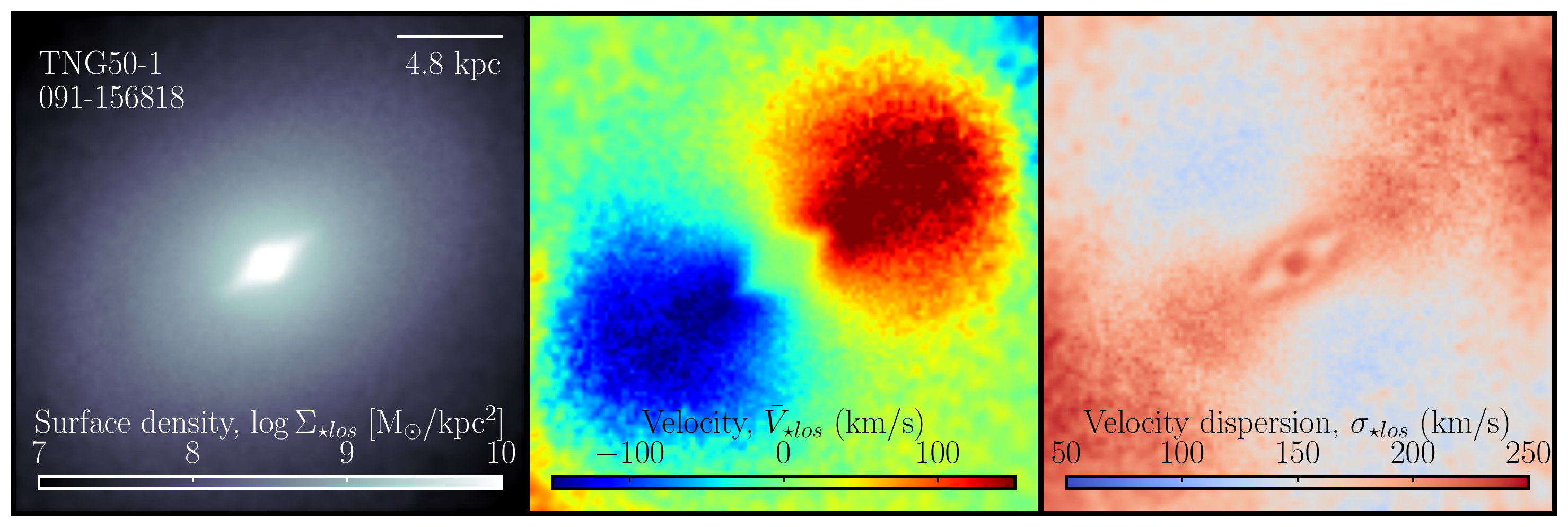}\\
	\vspace{-7pt}

	\includegraphics[width=0.78\linewidth]{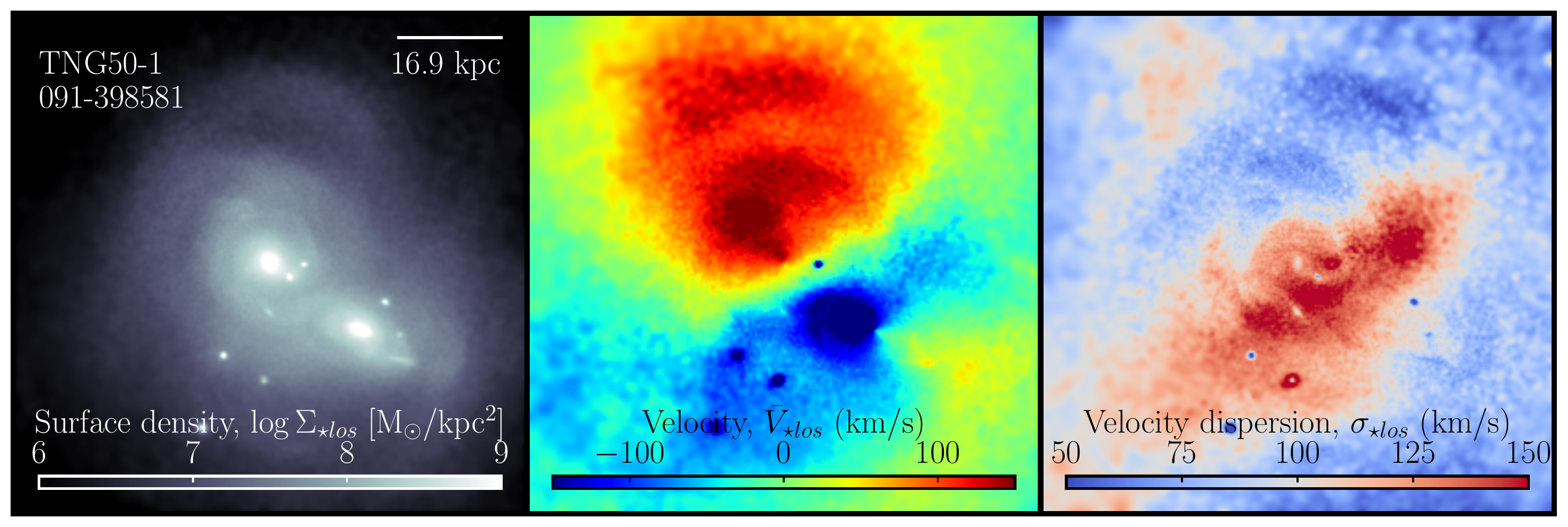}\\
	\vspace{-7pt}
	
	\includegraphics[width=0.78\linewidth]{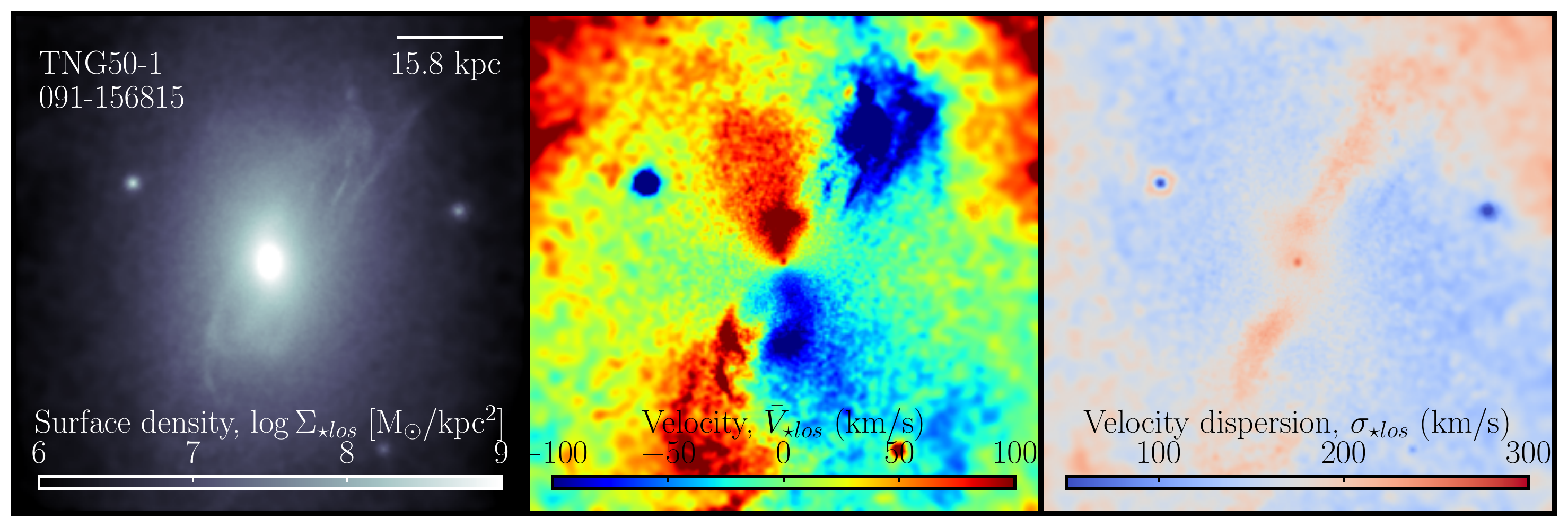}\\

   \caption[Idealized LOSVD moment maps]{Moments computed from idealized stellar LOSVD cubes for five TNG50 galaxies. From left to right, panels show the stellar surface density, velocity, and velocity dispersion, respectively. See Section \ref{sec:realifs_losvd} for further description. Labels on the upper left show the simulation ID and the \{snapshot\}-\{subfindID\} unique identifier. The scale is shown on the upper right. Each pixel is 200 pc.}
    \label{fig:realifs_ideal}
\end{figure*}

\subsection{LOSVD cubes}
\label{sec:realifs_losvd}

\rifs{} can be applied to any spatially resolved data cube as input for which non-spatial dimension is a light- or mass-weighted quantity: (1) wavelength; (2) frequency; (3) velocity; (4) stellar age; etc. The \emph{most realistic} approach to producing synthetic IFS data is to produce wavelength/frequency cubes using radiative transfer (e.g. \citealt{2006MNRAS.372....2J,2010MNRAS.403...17J,2011ApJS..196...22B,2013ARA&A..51...63S,2015A&C.....9...20C,2020A&C....3100381C}) -- from which all of the above properties may be extracted using the same methods used for real spectra and cubes. In this approach, each stellar particle is assigned a spectral energy distribution (SED) based on its mass, age, metallicity. Light packets can then be sampled from these SEDs and propagated through the gas and dust media (which can imprint their relative velocities in the emission and absorption features of the SED) to a camera which registers the spatial location and observer-frame wavelength/frequency of the light. The data format of the resulting idealized cubes are most similar to real IFS data but contain no observational limitations or nuisances. \rifs{} is primarily designed to incorporate realistic observational effects into such cubes. Spatially resolved spectra from this radiative transfer approach are most useful because all of the properties that may be extracted from IFS survey data can also be extracted from the synthetic data (potentially using the same data analysis pipeline). However, the higher fidelity from dusty, kinematic radiative transfer comes with a significantly increased computational overhead. Additionally, to subsequently extract kinematics, the gas and stellar LOSVDs in each spaxel must be derived from LOSVD-convolved emission line and stellar template fitting (e.g. \textsc{ppxf}, \citealt{2017MNRAS.466..798C}). This task goes well beyond the scope of testing and demonstrating applications of \rifs{}.

A less intensive approach is to use mass-weighted idealized LOSVD cubes made directly from the 6-dimensional particle data as input for \rifs{} (i.e. velocity cubes). In this approach, each spaxel of the resulting idealized cube contains LOSVD contributions from the gas or stellar particles it subtends. This avoids radiative transfer and spectral fitting, but results in data structures that are similar to spectral datacubes. Most importantly, incorporating the spatial response for an IFU design to idealized synthetic spectra or LOSVDs is programmatically equivalent. Consequently, for testing \rifs{}, we make idealized stellar LOSVD cubes for snapshot 91 ($z=0.1$) of the TNG50-1 simulations for \textsc{Subfind} subhalos \citep{2001MNRAS.328..726S} with stellar masses $\logMstar\geq10$. This stellar mass cut, which corresponds to a minimum subhalo stellar particle count $N_{\star}\sim120,000$ in TNG50-1, ensures that stellar kinematic structures (such as discs and spheroids) are reasonably resolved and that systematic differences between measured kinematic and morphological components are mitigated (e.g. \citealt{2010MNRAS.407L..41S, 2016MNRAS.459..467O, 2017MNRAS.467.2879B})\footnote{In particular, \cite{2017MNRAS.467.2879B} show that a systematic offset between morphological and kinematic measures of galaxy bulge fraction increases when $N_{\star}$ is reduced below $\sim10^5$ particles for galaxies in the Illustris simulation.}. These selection criteria yield a final sample of approximately 900 TNG50-1 galaxies.

The stellar LOSVD cube for each galaxy has dimensions $(N_x,N_y,N_v)$ where $N_x$ and $N_y$ are the spatial grid dimensions and $N_v$ is the number of velocity channels. The velocity axis ranges from $[-500,500]$ km s$^{-1}$ with resolution $4$ km s$^{-1}$ ($N_v=250$) with respect to an origin defined at the gravitational potential minimum of the galaxy. The FOV of the spatial axes are adaptively set to 6 times the physical half stellar mass radius of the galaxy, $R_{1/2}$ (radius containing 50\% of the galaxy stellar mass) with $200$ pc/pixel grid resolution. All friends-of-friends stellar particles from within $\sqrt{2}\times$FOV are considered (as opposed to subhalo particles only) to preserve satellites/companions and avoid numerical stripping effects \citep{2015MNRAS.449...49R,2018MNRAS.475.4066V,2020MNRAS.494.4969P}.The intensities in each velocity element have units of stellar mass, such that the summation over the velocity axis gives the total stellar mass within a spaxel. Lastly, each stellar particle in the simulation represents an unresolved stellar population of mass \Mstar{} $\approx8.5\times10^4$ \Msolar{}. Therefore, to avoid unrealistically discretized distributions of stellar mass in the cubes, a cubic spline smoothing kernel \citep{1992ARA&A..30..543M} is applied spatially to the data with an adaptive characteristic radius equal to the the distance to the 32$^{\mathrm{nd}}$ nearest neighbouring particle. LOSVD cubes are generated in four camera orientations along the arms of a tetrahedron that is randomly oriented with respect to the galaxy.

Figure \ref{fig:realifs_ideal} shows five examples of idealized stellar LOSVD moment maps (stellar surface density, velocity, velocity dispersion) derived from these cubes assuming single-component Gaussian LOSVDs in each spaxel\footnote{While the LOSVD in each spaxel may generally include contributions from multiple kinematic components which should be modelled simultaneously, a single-component Gaussian model is adopted for simplicity in our testing.} for which we compute unbiased estimators for the velocity and velocity dispersion \citet{2014A&C.....5....1R}:
\begin{align}
\Sigma_{\star} =&\frac{1}{p^2} \sum_{k=1}^{N_v} m_k \quad \text{where $p$ is the pixel scale (kpc);} \label{eq:moment0}\\ 
\bar{v} =&\frac{1}{V_1}\sum_{k=1}^{N_v}m_k v_k \quad \text{where $V_1=\sum_{k=1}^{N_v} m_k; \quad$ and}\label{eq:moment1}\\
\sigma^2 =& \frac{V_1}{V_1^2-V_2} \sum_{k=1}^{N_v} m_k (v_k - \bar{v})^2 \quad \text{where $V_2=\sum_{k=1}^{N_v} m^2_k$\label{eq:moment2}}
\end{align}
and $m_k$ is the mass in the $k^{\mathrm{th}}$ velocity element of a spaxel. The first rows of Figure \ref{fig:realifs_ideal} shows the brightest cluster galaxy (BCG) of TNG50's most massive group -- a giant elliptical with low rotation and high dispersion including stellar shell and stream substructures. The second row shows a fast-rotating spiral galaxy with a companion out-of-frame to the upper left. The third row shows a lenticular galaxy with clear rotation in its disk but also a relatively large high-dispersion bulge. The fourth row shows a spiral galaxy in a late-stage interaction -- also identifiable from the asymmetric structure in its velocity and velocity dispersion maps. Lastly, the fifth row shows a massive recent major merger remnant. 

\subsection{MaNGA}
\label{sec:realifs_manga}

Some tests with \rifs{} are carried out using MaNGA IFS data from Data Release $17$ (DR$17$) of the SDSS \citep{2019ApJS..240...23A}. The MaNGA survey \citep{2015ApJ...798....7B} has obtained observations for $\sim10,000$ nearby galaxies with $\logMstar\gtrsim9$ from the SDSS main galaxy sample \citep{2002AJ....124.1810S}. The survey uses a suite of custom-built fibre-based IFUs \citep{2015AJ....149...77D} fed into the Baryon Oscillation Spectroscopic Survey (BOSS) spectrographs \citep{2013AJ....146...32S} installed on the SDSS $2.5$ m telescope at Apache Point Observatory \citep{2006AJ....131.2332G}. Each IFU comprises a number of 2 arscecond fibres bundled in a hexagonal pattern. 

The IFUs cover a range of fibre numbers and corresponding observational footprints. The IFUs used for galaxy observations are bundles of $19,\;37,\;61,\;91,\;\mathrm{and}\;127$ individual fibres with respective FOV diameters of $12.5,\;17.5,\;22.5,\;27.5,\;\mathrm{and}\;32.5$ arcseconds. The IFU design for a given galaxy is selected to cover out to $1.5$ effective radii for $63\%$ of the full MaNGA sample (called Primary and Colour-Enhanced samples) and $2.5$ effective radii for $37\%$ (Secondary sample) \citep{2017AJ....154...86W}. A more detailed description of the target selection follows in Section \ref{sec:realifs_survey} from which synthetic observation statistics are drawn (such as redshift and atmospheric seeing). 

Each MaNGA observation comprises between $2$ and $7$ sets of three dithered exposures with a given IFU to (1) fill gaps between fibres in each exposure's observational footprint, (2) increase signal, and (3) enable outlier rejection \citep{2015AJ....150...19L}. DR17 is the final MaNGA data release and includes raw fibre spectra as well as intermediate and final data products for the full $\sim10,000$ galaxy sample: including calibrated and sky-subtracted fibre spectra and spatially reconstructed data cubes. The testing performed in Section \ref{sec:realifs_accuracy} make use of both the individual calibrated fibre spectra and the final data cubes. 

\section{Methods}
\label{sec:realifs_methods}

This section describes the core functionalities of \rifs{} for reproducing the instrumental responses and corresponding covariances of IFS instruments in data from hydrodynamical simulations. The core functionalities are: (1) generating fibre patterns and dithering strategies; (2) integrating the intensities from spatially discretized input datacube spaxels into arbitrary fibres; and (3) the distribution of (potentially) irregularly sampled fibre intensity measurements onto regularly sampled Cartesian grids (output cubes). These main components are geared toward IFUs which collect data in irregularly sampled fibres and subsequently spatially reconstruct onto a regular Cartesian grid. Consequently, emulating the instrumental sampling mechanics of image slicer and lenslet array IFUs does not require these steps and can be handled with the code's more trivial components for dealing with regular Cartesian data (e.g. convolution of atmospheric and instrumental point spread functions).

\subsection{Instrumental designs and observing strategies}
\label{sec:realifs_ifus}

\begin{figure}
\centering
	\includegraphics[width=\linewidth]{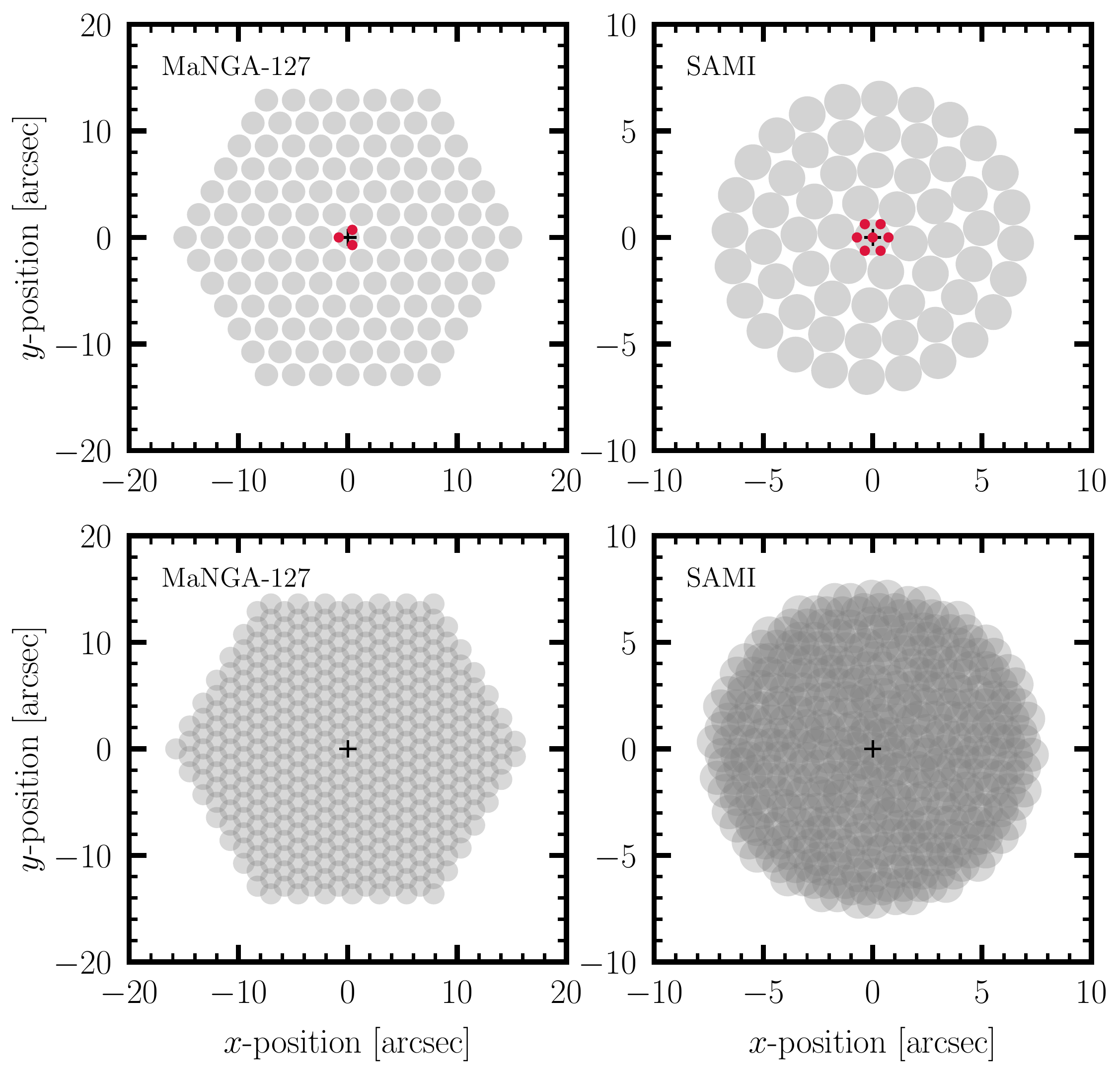}
    \caption[Instrumental designs and observing strategies]{MaNGA (upper left) and SAMI (upper right) IFU designs and respective observing strategies (lower panels) emulated by \rifs{}. Individual fibre cores are marked by light-grey shaded circles. Black crosses mark the centre of the FOV (typically galaxy barycenter). Upper panels show the fibre pattern for a single exposure made by an IFU from each instrument. The IFU centroid positions of a complete set of dithered exposures are indicated by red markers in the upper panels -- three for MaNGA, seven for SAMI. The dithered exposures are shown in the lower panels in which increased shading shows increased fibre overlap.}
    \label{fig:realifs_ifus}
\end{figure}

The designs of the observing units for modern fibre-based IFS instruments vary significantly. These designs are generally optimized to a set of science objectives and the boundaries set by (1) the precision to which new components can be fabricated, (2) existing components (e.g. telescopes and spectrographs), and (3) cost. Furthermore, the observing strategies for these instruments also vary greatly. Consequently, \rifs{} must be sufficiently flexible that existing \emph{and anticipated} fibre-based IFU designs and observing strategies can be emulated. 

These requirements are accommodated by allowing a user to provide an arbitrary array of on-sky fibre positions  and corresponding core diameters (in arcseconds). In general, these fibres are free to overlap -- as it is assumed that consecutive observations of the same piece of sky are possible. It is up to the user to ensure that their synthetic IFU design and observing strategies are feasible. However, since the packing and positioning of fibres in real IFUs are physically limited by the diameters of fibre cores and their surrounding buffer/cladding, it is useful to adopt a single IFU design and perform repeated measurements with an offset pattern (\emph{dithering}). The purpose of dithering is to improve (1) the fill-factor of the FOV that is subtended by fibres and (2) the signal-to-noise in regions where fibres from consecutive observations overlap. Consequently, in addition to an arbitrarily long list of potentially overlapping fibres, a \rifs{} user can provide an array of fibre positions corresponding to a single IFU design and the $[x_1,...,x_n]$ and $[y_1,...,y_n]$ offsets for each of the $n$ dithered exposures. \rifs{} also includes tailored modules which are hard-coded to emulated the current state-of-the-art fibre-based IFU designs and observing strategies of MaNGA and SAMI using information gathered from the corresponding technical papers \citep{2015AJ....149...77D,2015AJ....150...19L,2015MNRAS.447.2857B,2015MNRAS.446.1551S}\footnote{The fibre positions in all MaNGA IFU designs and dithering offsets were derived analytically using the information from the survey strategy paper \citep{2015AJ....150...19L}. The SAMI fibre bundle positions were generously provided by Nic Scott on behalf of the SAMI Survey team.}. These tailored modules allow a user to automatically set up SAMI and MaNGA IFU designs and strategies (including \emph{all} current MaNGA IFU sizes). 

Figure \ref{fig:realifs_ifus} shows MaNGA and SAMI IFU designs and observing strategies as set-up by \rifs{}. These designs are ``idealized'' in the sense that they do not incorporate random or systematic errors involved in the manufacture of the IFUs, for example. The black crosses mark the centre of the FOV (all panels) and the centre of the IFUs (upper panels). The upper left panel shows the MaNGA 127-fibre IFU design. Shaded light-grey circles show the individual fibre cores of the IFU: $2$ arcseconds ($120\;\mu$m) in diameter. Each MaNGA fibre core is surrounded by a buffer and cladding that that is $15\;\mu$m in thickness -- bringing the total diameter of each fibre to $2.5$ arcseconds ($150\;\mu$m). The diameter of the full MaNGA 127-fibre IFU is around $32.5$ arsceconds. The right panel shows the SAMI 61-fibre IFU design. SAMI fibre cores are 1.6 arcseconds ($105\;\mu$m) in diameter with a buffer and cladding thinned to 0.076 arcseconds ($5\;\mu$m) for denser packing. The diameter of the full SAMI 61-fibre IFU is around 14.7 arcseconds.

The MaNGA and SAMI surveys employ different dithering strategies that are optimized for their respective IFU designs. The small red markers in the upper panels of Figure \ref{fig:realifs_ifus} show the centres of each dithered IFU exposure for a given MaNGA and SAMI target. For MaNGA (upper and lower left panels), the IFU in each of three dithered exposures is centred at the vertex of an equilateral triangle (red markers) whose centroid is the centre of the FOV (black cross, typically the target galaxy's barycentre). MaNGA observations consist of several repeats of this three-point dither pattern to increase signal-to-noise and enable outlier rejection. The SAMI survey (right panels) uses an optimized seven-point dither pattern comprising the vertices and centroid of a hexagon for which the radial distances between the centroid and vertices are $0.72$ arcseconds ($45\%$ of the fibre core diameter, \citealt{2015MNRAS.446.1551S}). 

With the on-sky positions and diameters of each fibre for a chosen observing strategy, fibre measurements of the synthetic datacubes can be made. In the next section, we describe the method used by \rifs{} to spatially integrate synthetic datacube intensities within fibres of arbitrary positions and apertures.

\subsection{Fibre observations of spatially discretized data}
\label{sec:realifs_fibreobserve}

\begin{figure}
\centering
	\includegraphics[width=\linewidth]{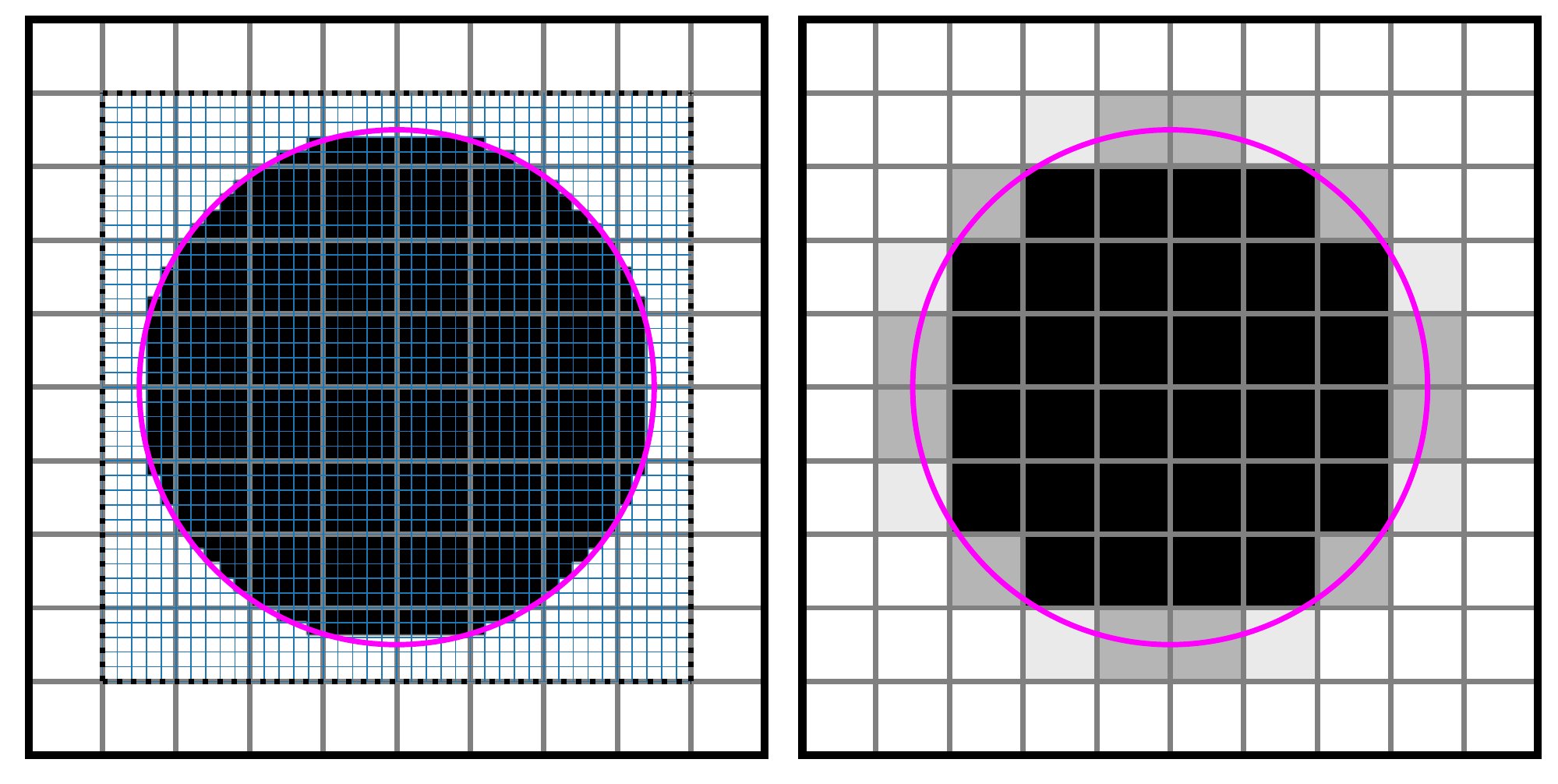}
    \caption[Fibres and spatially discretized data]{A simple algorithm is used to weight and integrate the contributions from spatially discretized datacube spaxels (grid of grey lines) within an arbitrary fibre (magenta). First, as illustrated in the left panel, a rectangular grid (black, dotted outline of blue sub-grid) is defined such that all spaxels which are completely or partially enclosed by the fibre aperture are within. The grid is then refined into sub-spaxels (blue squares) such that \emph{at least} 100 sub-spaxels (or original spaxels) subtend the fibre diameter. For visibility, each original spaxel in the figure is refined simply by a factor of five. The fraction of sub-spaxels in each original spaxel whose centres are within the fibre aperture is taken as the weight of that original spaxel in the spatial integration. The right panel shows the resulting weight-map in greyscale.}
    \label{fig:realifs_grid}
\end{figure}

The input for \rifs{} is a discretized datacube with a regular Cartesian spatial grid. For a given redshift, the physical resolution of the spatial grid \emph{should} sample the physical scale subtended by each fibre at the Nyquist rate or greater -- but this is not enforced. The non-spatial channels are intensity (weight) maps in bins of a light- or mass-weighted quantity (in our demonstration case, mass-weighted stellar velocity). To measure the intensities registered within an arbitrary fibre, an algorithm for spatially integrating datacube spaxels within the fibre aperture is needed.

Figure \ref{fig:realifs_grid} illustrates the simple algorithm used by \rifs{} to integrate spatially discretized input datacube spaxels in arbitrary fibre apertures. The algorithm first isolates a rectangular set of spaxels around a given fibre to restrict weight calculations to a relevant subset of original spaxels. These spaxels are then further refined spatially by a factor which guarantees that there are at least 100 spatial elements along the diameter of the fibre (for visibility, the original spaxels in Figure \ref{fig:realifs_grid} are refined only by a factor of five). The choice of 100 as the minimum number of spatial elements subtending the fibre diameter guarantees sub-percent accuracies in the weight and subsequent spatial integration calculations. The fraction of sub-spaxels of a given original spaxel whose centres are within the fibre aperture is taken as the weight of that original spaxel in the spatial integration. The resulting weight-map, $w[i,j]$ (right panel), is then applied in a weighted sum to the intensity map corresponding to each non-spatial channel, $f[i,j]$, to produce the scalar intensity registered by the fibre in that channel, $F$:
\begin{align}
\label{eq:realifs_fibreobserve}
F = \sum_{i=1}^{N_y}\sum_{j=1}^{N_x} \;w[i,j]\; f[i,j]
\end{align}

The outcome after applying the fibre's weight-map to each channel is then a one-dimensional fibre array of length $N_{\mathrm{ch}}$, the number of channels. After obtaining $N_{\mathrm{fib}}$ fibre measurements of the data, the result is a two-dimensional $N_{\mathrm{fib}} \times N_{\mathrm{ch}}$ matrix in which each row is the one-dimensional intensity distribution in the non-spatial axis registered by a given fibre. If the non-spatial axis is wavelength, then the data resulting from this step are analogous to the row-stacked spectra (RSS) data products from observational IFS surveys (e.g. see Section \ref{sec:realifs_accuracy}).

Conveniently, the algorithm used here is an inversion of the ``drizzle'' algorithm \citep{2002PASP..114..144F} used by various fibre-based IFS instruments to distribute spatially irregular intensity measurements onto regular Cartesian grids (e.g. SAMI: \citealt{2015MNRAS.446.1551S}). Elements of this same algorithm are incorporated into the drizzling modules of \rifs{} that are specific to instruments which use this data reduction strategy. 

\subsection{Spatial reconstruction of fibre measurements}
\label{sec:realifs_grid}

With the stacked measurements for each fibre, the final step for \rifs{} is to spatially redistribute these measurements onto a regular Cartesian grid of output spaxels. Mirroring the variety in IFU designs and dithering strategies, there are several algorithms that can be employed to spatially reconstruct the flux from non-Cartesian sampled fibre measurements. 

First, there is the simple inverse of the algorithm employed by \rifs{} to make fibre measurements as detailed in Section \ref{sec:realifs_fibreobserve}. This method is adopted by the SAMI survey. The SAMI pipeline uses a substantial modification to the standard drizzling approach in which the intensities from a given fibre are re-distributed within spaxels in an aperture that is half the size of the actual fibre core. This strategy, afforded by SAMI's large number of dithered exposures, enables improved preservation of spatial resolution \citep{2015MNRAS.446.1551S}. Another is the modified (flux-conserving) Shepard approach adopted by the CALIFA and MaNGA data reduction pipelines (DRPs, \citep{2012A&A...538A...8S,2016AJ....152...83L}. \rifs{} provides both the drizzling and modified Shepard algorithms as options for spatial reconstruction of the fibre measurements and each is described below.

\subsubsection{Drizzling algorithm}
\label{sec:drizzling}

The intensity distributed to a given output spaxel (for a given channel) by a given fibre is determined through a spatial weighting scheme as illustrated in Figure \ref{fig:realifs_grid}. Spaxels completely contained within the $k^{\mathrm{th}}$ fibre's aperture are assigned a non-normalized weight, $w[k,i,j] = 1$. The non-normalized weights on spaxels that partially overlap with the aperture are computed from their fractional overlapping area with the aperture. The weight assigned to each $[i,j]$ spaxel from each $[k]$ fibre is then:
\begin{align}
\label{eq:realifs_normalize}
W[k,i,j] = \alpha[k] \frac{w[k,i,j]}{\sum_{k=1}^{N_{\mathrm{fib}}} w[k,i,j]} 
\end{align}
where
\begin{align}
\label{eq:realifs_normalization}
\alpha^{-1}[k] = \sum_{i=1}^{N_y} \sum_{j=1}^{N_x} w[k,i,j]
\end{align}
Each fibre core on the IFU (which may or may not be of the same radius, $r_{\mathrm{core}}[k]$) has an associated flux scaling factor, $\alpha$, corresponding to the area, in output spaxels, subtended by the fibre core, $\alpha^{-1}[k] = \pi r_{\mathrm{core}}^2[k]$. This combination of fibre normalization and flux scaling guarantee that calibrated flux is conserved within the limits of the spatial sampling. With these normalizations, the calibrated flux in a given output channel, $I[i,j]$, is:
\begin{align}
I[i,j] = \sum_{k=1}^{N_{\mathrm{fib}}} W[k,i,j] \; F[k]
\end{align}
Alternatively, to conserve intensity only, such that the sum of the flux in an output channel is equal to the sum of the flux in that channel from each fibre, the normalization term in the denominator of Eq. \ref{eq:realifs_normalize} can be ignored. \rifs{} provides pre-normalization and pre-scaling weights in addition to the output cubes to make it straight-forward to conserve total integrated intensity rather than calibrated flux, if desired. In either case, the weight maps are valuable for computing the corresponding variance maps for each output channel: 
\begin{align}
\label{realifs_variance}
V[i,j] = \sum_{k=1}^{N_{\mathrm{fib}}} W^2[k,i,j] \; V[k]
\end{align}

The specific algorithm used by the SAMI survey is a slight modification of the drizzling algorithm described above in which the ``drop radius'' of each fibre's flux on the output grid is 50\% of the original fibre size (see \citealt{2015MNRAS.446.1551S}). This adaptation is easily implemented in \rifs{}. First, the size of the fibre aperture in the output grid can be set manually to the desired drop radius. Second, provided all fibre cores have the same diameter, flux conservation can be achieved by multiplying the output cube by the factor, $\zeta^2 = (r_{\mathrm{drop}}/r_{\mathrm{core}})^2$, which derives from the relative areas of the drop aperture and ``true'' core aperture. If fibres are not all of the same diameter, $\zeta$ is specific to each fibre and must be integrated into the weights. In either case, $\zeta$ must be incorporated into the variance maps.

\subsubsection{Flux-conserving  Shepard algorithm}

Modifications of the Shepard algorithm \citep{10.1145/800186.810616} are used by the CALIFA and MaNGA data reduction pipelines (DRPs, \citealt{2012A&A...538A...8S,2015AJ....150...19L}). The non-normalized flux contribution from each fibre to each regularly spaced output spaxel is mapped with a two-dimensional Gaussian distribution:
\begin{align}
\label{eq:gaussian_weights}
w[k,i,j] = q[k,i,j] \exp \left( \frac{- r^2[k,i,j]}{ 2\sigma^2}\right)
\end{align}
where $r[k,i,j]$ is the radial distance from the $k^{\mathrm{th}}$ fibre core centroid to the centroid of output spaxel $[i,j]$ and $\sigma$ is an adopted standard deviation for the Gaussian distribution (taken to be 0.7 arcseconds or 1.4 output grid spaxels in the MaNGA DRP). Because the Gaussian distribution is non-zero everywhere, a characteristic truncation radius, $r_{\mathrm{lim}}$, can be defined, beyond which $w[k,i,j]=0$. This truncation is represented by the $q[k,i,j]$ term -- which is a binary mask equal to 1 where $r[k,i,j] \leq r_{\mathrm{lim}}$ and 0 beyond. It allows the conservation of total fiber flux within a finite spatial domain in the output grid. In MaNGA, $r_{\mathrm{lim}}$ is taken to be 1.6 arcseconds or 3.2 output spaxels. The effect of $r_{\mathrm{lim}}$ on the correlation coefficients between spaxels is nicely illustrated in Figure 7 of \citet{2019AJ....158..231W}. 

The weights are normalized to conserve flux. This normalization follows Eq. \ref{eq:realifs_normalize} using the initial non-normalized Gaussian weights, $w[k,i,j]$, from Eq. \ref{eq:gaussian_weights}. However, the non-normalized Gaussian weights for a given fibre do not sum to the fibre area in output spaxels as they did in Eq. \ref{eq:realifs_normalization}. Therefore, the $\alpha$ term is computed manually as $\alpha^{-1}[k] = \pi r_{\mathrm{core}}^2[k]$ where $r_{\mathrm{core}}[k]$ is the radius, in output spaxels, of the $k^{\mathrm{th}}$ fibre core. However, as with any algorithm for redistributing fibre fluxes onto Cartesian grids, perfect flux conservation is only guaranteed with complete (and extended) spatial sampling of the original data by the fibres.

For both spatial reconstruction methods used by \rifs{}, it is assumed that the astrometric positions of the fibres are the same for all channels (e.g. wavelength). We make note of wavelength-dependent astrometric effects on the effective positions of fibres in our reconstruction tests using real MaNGA fibre data in Section \ref{sec:realifs_accuracy}. 

\section{Test results and demonstrations}
\label{sec:realifs_results}
This section comprises several demonstrations with \rifs{} using the MaNGA and TNG50 data outlined in Section \ref{sec:realifs_data}. First, we demonstrate the precision with which \rifs{} specifically emulates the spatial reconstruction component of the MaNGA DRP using real MaNGA data (Section \ref{sec:realifs_accuracy}) and highlight some of the higher-order limitations of \rifs{}. Then, using similar targeting selection to MaNGA, we construct a synthetic MaNGA stellar kinematic survey using similar target selection and IFU assignment strategies to those used by the MaNGA survey, additionally incorporating atmospheric point-spread functions and spectral line-spread functions (Section \ref{sec:realifs_survey}).

\subsection{Precise emulation of MaNGA data reduction}
\label{sec:realifs_accuracy}

\begin{figure*}
\centering
	\includegraphics[width=\linewidth]{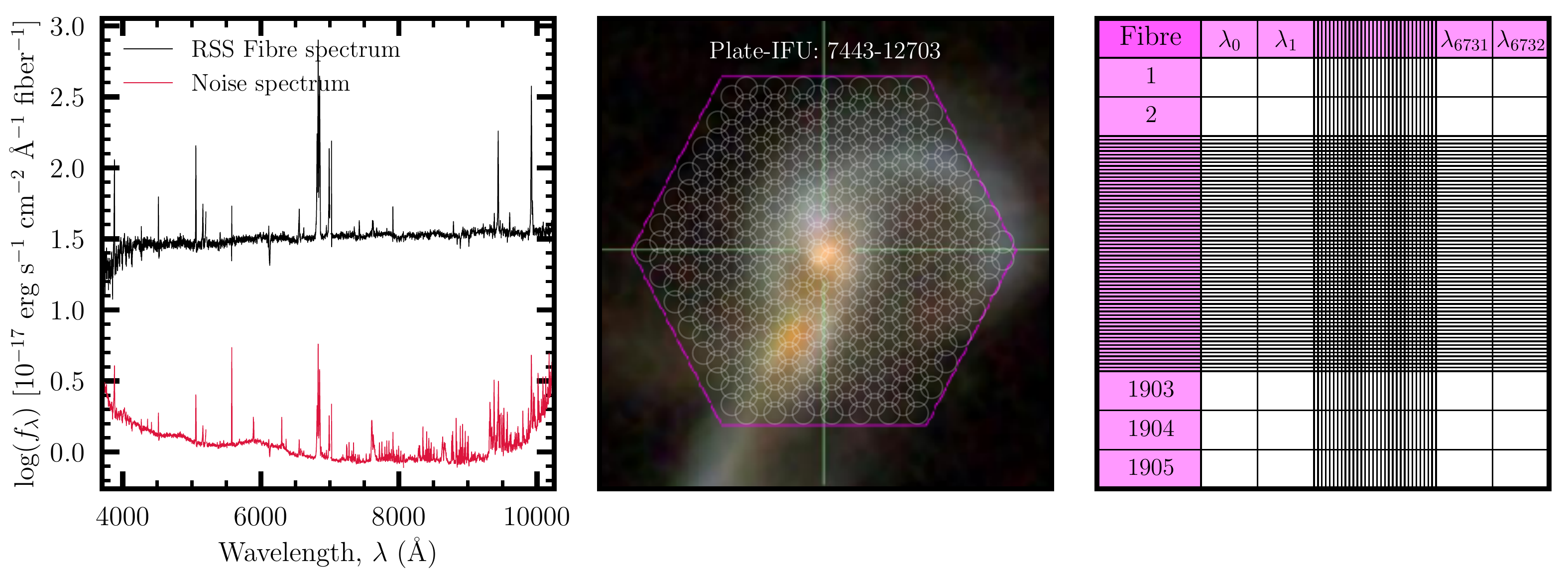}
    \caption[Row-stacked spectra data]{Infographic on row stacked spectra (RSS) data structure format. The left panel shows a spectrum and corresponding noise spectrum from the RSS file for MaNGA Plate-IFU target 7443-12703 produced by the MaNGA DRP. The spectrum is taken from near the centre of the IFU footprint. The central panel shows the $gri$ image of MaNGA Plate-IFU target 7443-12703 overlaid with three dithered 127-fibre IFUs. Five repeated observations with this three-point dither pattern were used for a total of $5\times3\times127=1905$ fibre spectra. The right panel shows one of the data structures contained within the RSS file: the calibrated spectrum from each fibre core. These data are used in the reconstruction precision tests in Section \ref{sec:realifs_accuracy}. RSS format is also an intermediate output from \rifs{}.}
    \label{fig:realifs_rss}
\end{figure*}

\begin{figure}
\centering
	\includegraphics[width=0.9\linewidth]{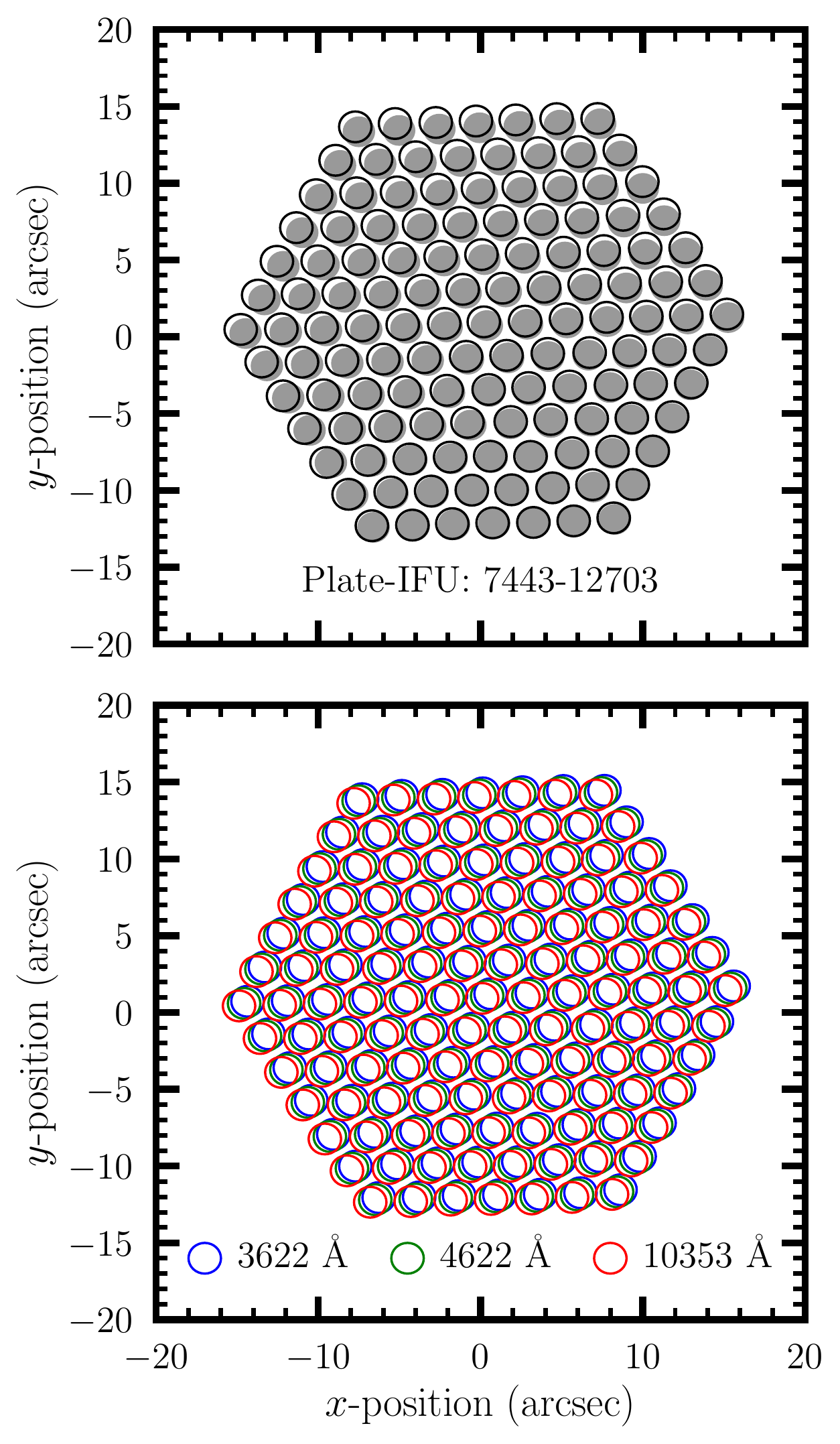}
    \caption[The effect of astrometric biases]{The effect of astrometric corrections on the wavelength-dependent  effective positions of fibres relative to their intended pattern. The upper panel shows the astrometric offsets between the idealized IFUs (grey filled circles) and the median effective fibre positions over all wavelength channels for a single MaNGA 7443-12703 exposure (black open circles). The lower panel specifically shows the astrometrically corrected positions of fibres for three wavelength channels for the same exposure -- accounting for differential atmospheric refraction. The data reduction pipelines of observational IFS campaigns generally correct for these effects with $\sim 0.1$ arcseconds precision \citep{2015MNRAS.446.1551S,2016AJ....152...83L}.}
    \label{fig:realifs_astrometry}
\end{figure}

The precision with which \rifs{} can reproduce data reduction of real fibre measurements can be tested explicitly using real observations. IFS surveys typically generate two main data products: (1) row-stacked spectra (RSS) files and (2) spectral data cubes. The RSS files contain the calibrated spectrum from each fibre core and a wealth of ancillary data (including astrometrically correct fibre positions). The RSS files are therefore a vital intermediate data product between the raw spectra and the output cubes. Figure \ref{fig:realifs_rss} shows an infographic of a MaNGA RSS file.

The RSS files provide the information required to spatially reconstruct the IFU fibre measurements onto a regular Cartesian grid. In the MaNGA DRP (at least), the output wavelength cubes (e.g. LINCUBE) are the direct results of spatial reconstruction of the spectra in the RSS files. Consequently, one straight-forward test is to examine whether \rifs{} reproduces this spatial reconstruction component using real fibre data from the RSS files. We carry out this test with MaNGA Plate-IFU 7443-12703 (the same target as from Figure \ref{fig:realifs_rss}). This target was chosen for testing arbitrarily from the 127-fibre IFU MaNGA targets and is conveniently a visually stunning interacting pair that belongs to the MaNGA galaxy pair and binary AGN samples of \citet{2018ApJ...856...93F}.

For our spatial reconstruction of the MaNGA RSS data, we use the associated wavelength-dependent astrometric corrections to the effective fibre positions. These astrometric corrections account for offsets in the effective positions of fibres compared to their intended positions. Figure \ref{fig:realifs_astrometry} illustrates why these corrections are important for our spatial reconstruction test. The upper panel shows the median effective positions of fibres, estimated over all wavelength channels, for a single MaNGA exposure (black open circles) relative to the ideal positions used by \rifs{} (i.e. which assumes no wavelength-dependence or manufacture imprecision in fibre positions). The lower panel sheds light on the specific role of differential atmospheric refraction (DAR, e.g. \citealt{1982PASP...94..715F}). 

DAR is generated by the increased refractive index of the atmosphere for shorter wavelengths than longer wavelengths -- an effect which is exacerbated with increasing airmasses (i.e. the greater the angle between the target and the zenith, the greater the effects of DAR). If the airmass is known, then DAR can be reasonably accounted for in the effective astrometric positions of the fibres in each wavelength channel. Taken together, the wavelength-dependent astrometry in fibre position offsets is corrected to $\sim 0.1$ arcsecond precision by the SAMI and MaNGA DRPs, for example \citep{2015MNRAS.446.1551S,2016AJ....152...83L}. This precision error is dwarfed by the $2-2.5$ arcsecond effective spatial resolution of these IFS surveys -- but can nonetheless be easily introduced in \rifs{} using a 0.1 arcsecond Gaussian random error in the output fibre positions for the spatial reconstruction step.  

\begin{figure*}
\centering
	\includegraphics[width=\linewidth]{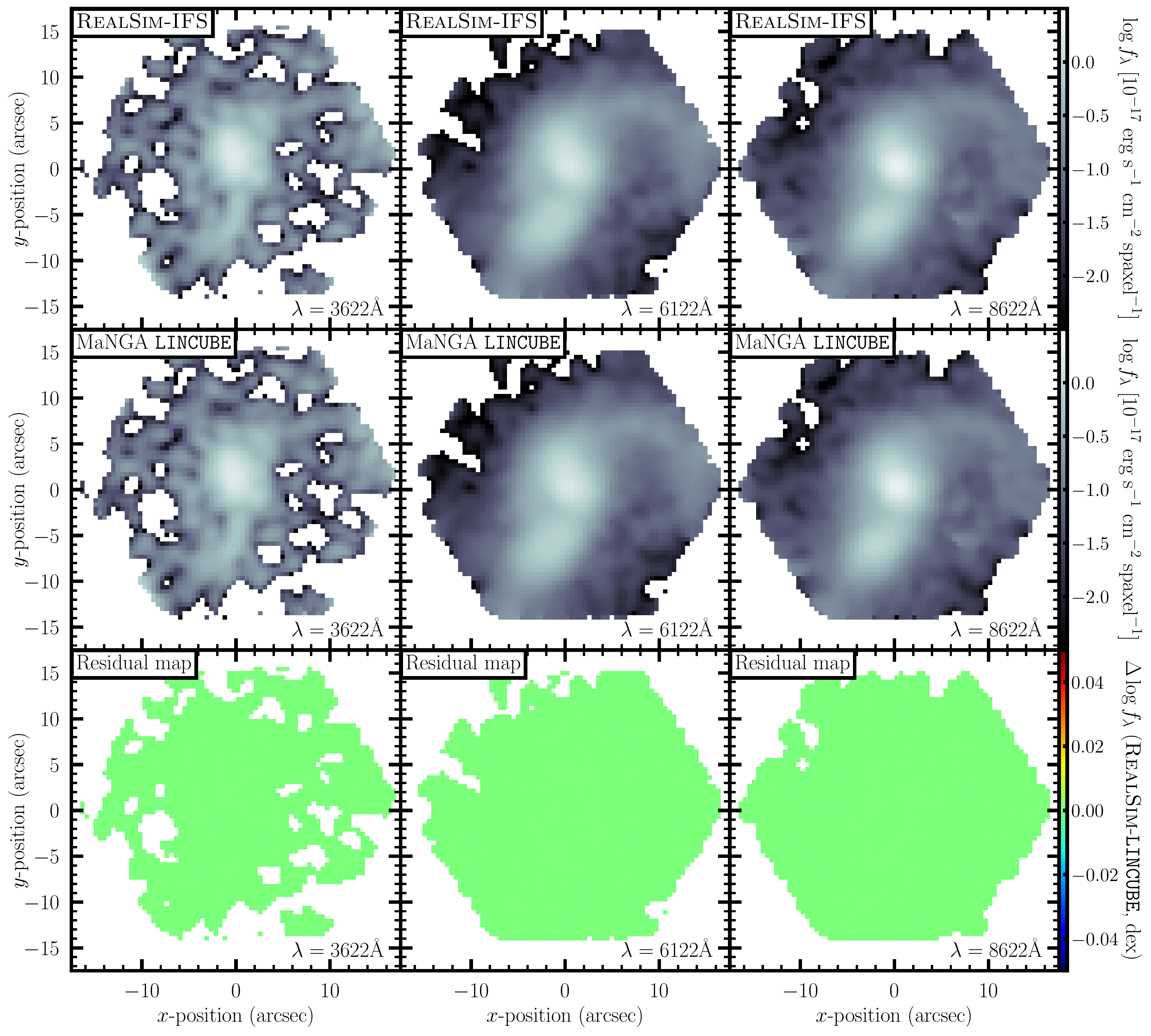}
    \caption[Flux reconstruction]{Spatial reconstruction of MaNGA Plate-IFU 7443-12703 flux cube from the RSS fibre data in three wavelength channels using \rifs{} (upper panels) and by the MaNGA DRP (\texttt{LINCUBE}, middle panels). The lower panels show the residual maps (\rifs{}$-$\texttt{LINCUBE}). Using the wavelength dependent fibre astrometry (accounting for differential atmospheric refraction) and masking bad data with the RSS variance spectra with \rifs{}, the channel fluxes and spatial distributions are exactly reproduced.}
    \label{fig:realifs_slices}
\end{figure*}

\begin{figure*}
\centering
	\includegraphics[width=\linewidth]{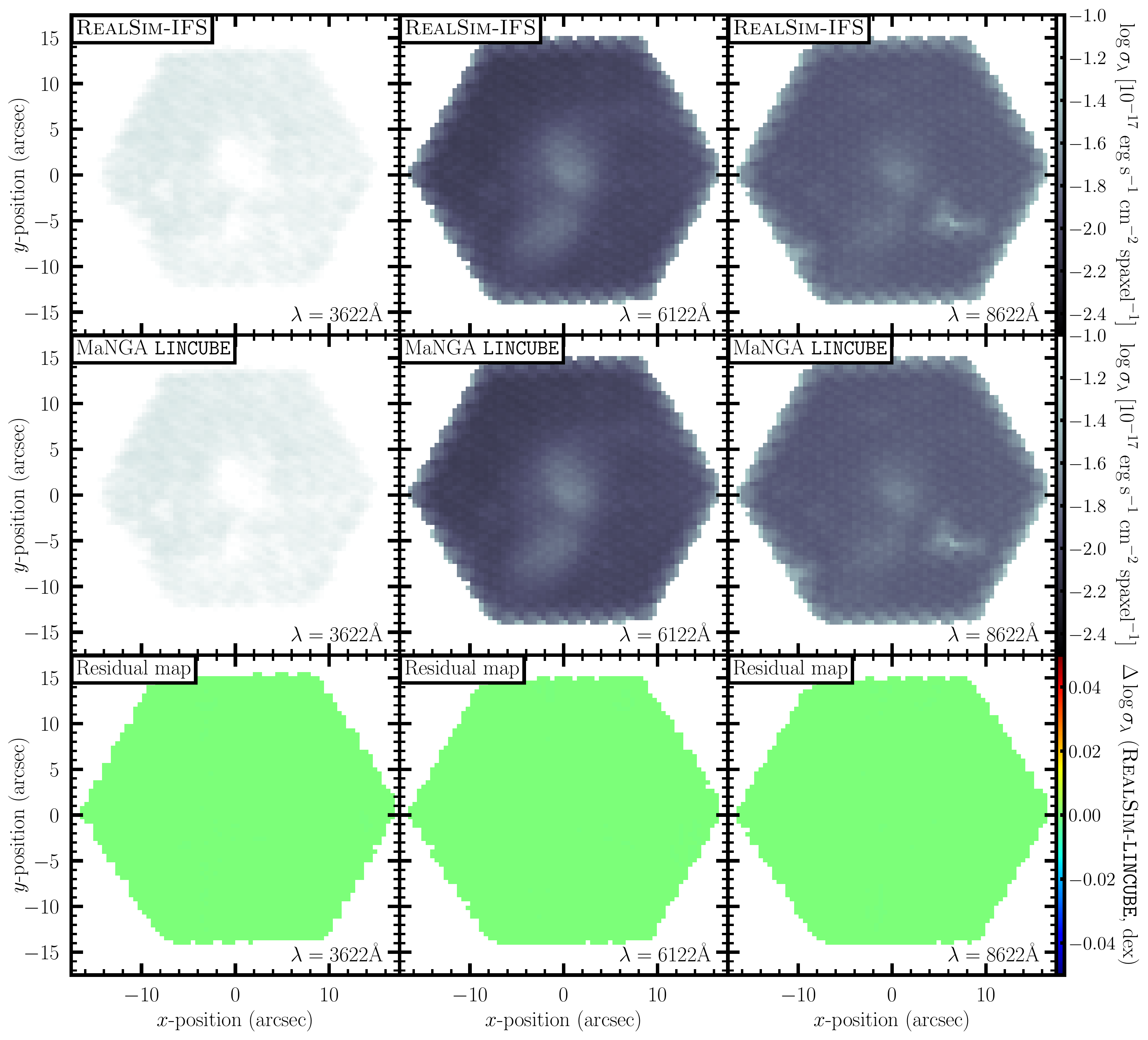}
    \caption[Variance reconstruction]{Similar to Figure \ref{fig:realifs_slices} but comparing of the noise cubes reconstructed by \rifs{} from MaNGA Plate-IFU 7443-12703 noise spectra to those produced by the MaNGA DRP. The variance propagation by the MaNGA data reduction pipeline is exactly reproduced by \rifs{}.}
    \label{fig:variance_slices}
\end{figure*}

A simple wrapper for \rifs{} was used to reconstruct the flux independently in each wavelength channel using the corresponding channel-specific astrometric corrections. The upper row of Figure \ref{fig:realifs_slices} shows the \rifs{} reconstructed flux in three wavelength channels, $(3622, 6122, 8622)$\AA, using the fibre spectra for MaNGA Plate-IFU 7443-12703. The variance maps from the RSS files were used to mask data with arbitrarily high variance (\texttt{IVAR}$=0$)\footnote{Masking of data with exceptionally high variance eliminates the signatures of artifacts such as cosmic rays, dead fibre elements in the IFUs, and dead CCD pixels in the spectrographs.}. The middle row shows the flux in the same channels from the MaNGA LINCUBE for this target. The lower row shows offsets between the \rifs{} and MaNGA DRP flux reconstructions in each channel. \rifs{} reconstructs the specific fluxes for MaNGA Plate-IFU 7443-12703 with an exceptional median precision of $\Delta \log f_{\lambda} \lesssim 10^{-8}$ dex in each channel. Additionally, the footprints of \emph{good} spaxels are identical. 

\rifs{} provides built-in support for the propagation of variances following Equation \ref{realifs_variance} for both the Drizzling and Modified Shepard Algorithms. Using this functionality, the variances in each wavelength channel were also reconstructed from the RSS noise spectra. Figure \ref{fig:variance_slices} compares the reconstructed noise maps, $\sigma_{\lambda}$, for three wavelength elements using \rifs{} (upper panels) and from the MaNGA cubes (middle panels). The residual maps in the lower panels show that the reconstruction of the MaNGA cube variance maps is exactly reproduced by \rifs{}. 

In summary, \rifs{} can be used to precisely emulate the spatial flux reconstruction components of modern IFS surveys. Consequently, a high level of realism can be achieved for synthetic data produced by \rifs{} -- including the distributions of fluxes, variances, and spatial covariances. Satisfaction of the spatial reconstruction test provides a strong foundation for using \rifs{} for its intended purpose: creation of survey-realistic \emph{synthetic} IFS observations.

\subsection{Synthetic MaNGA kinematic observations}
\label{sec:realifs_survey}

In this section, we describe an application of \rifs{} to create a synthetic MaNGA survey for a sample of $893\times4=3572$ galaxy lines-of-sight with $\logMstar\geq10$ from the TNG50 cosmological hydrodynamical simulation -- specifically focusing on stellar kinematics. The four sightlines increase the sample volume without redundancy and enable the evaluation of parameter sensitivities to galaxy orientation. The input data for this synthetic survey are the idealized stellar kinematic cubes described in Sections \ref{sec:realifs_losvd}. To incorporate realism into these synthetic data, the first step is to \emph{roughly} emulate the MaNGA survey selection and observing strategies for the simulated galaxy sample. This step includes the assignment of:
\begin{enumerate}
\item Redshifts at which to mock-observe each galaxy: setting angular size and physical-to-angular scale.
\item Atmospheric seeing conditions to each synthetic observation: setting the pre-instrumental spatial resolution.
\item Appropriate IFUs to each galaxy based on galaxy angular size and the desired relative field-of-view.
\item Instrumental line-spread functions into the line-of-sight velocity distributions in each cube.
\end{enumerate}  
Our procedures for emulating the four components are described in the subsections which follow. We do not emulate the complete MaNGA selection function in our synthetic observations. Instead, we produce synthetic MaNGA stellar kinematic observations for every TNG50 galaxy in our sample and take steps to ensure that their mock-observed redshifts, seeing, IFU assignment, and velocity resolution are commensurate with MaNGA observations. We do not make cuts in our TNG50 sample to simultaneously match the redshift-luminosity distribution of MaNGA, for example, as highlighted in the next subsection.

\subsubsection{Redshift selection}\label{sec:realifs_redshifts}

\begin{figure}
	\includegraphics[width=\linewidth]{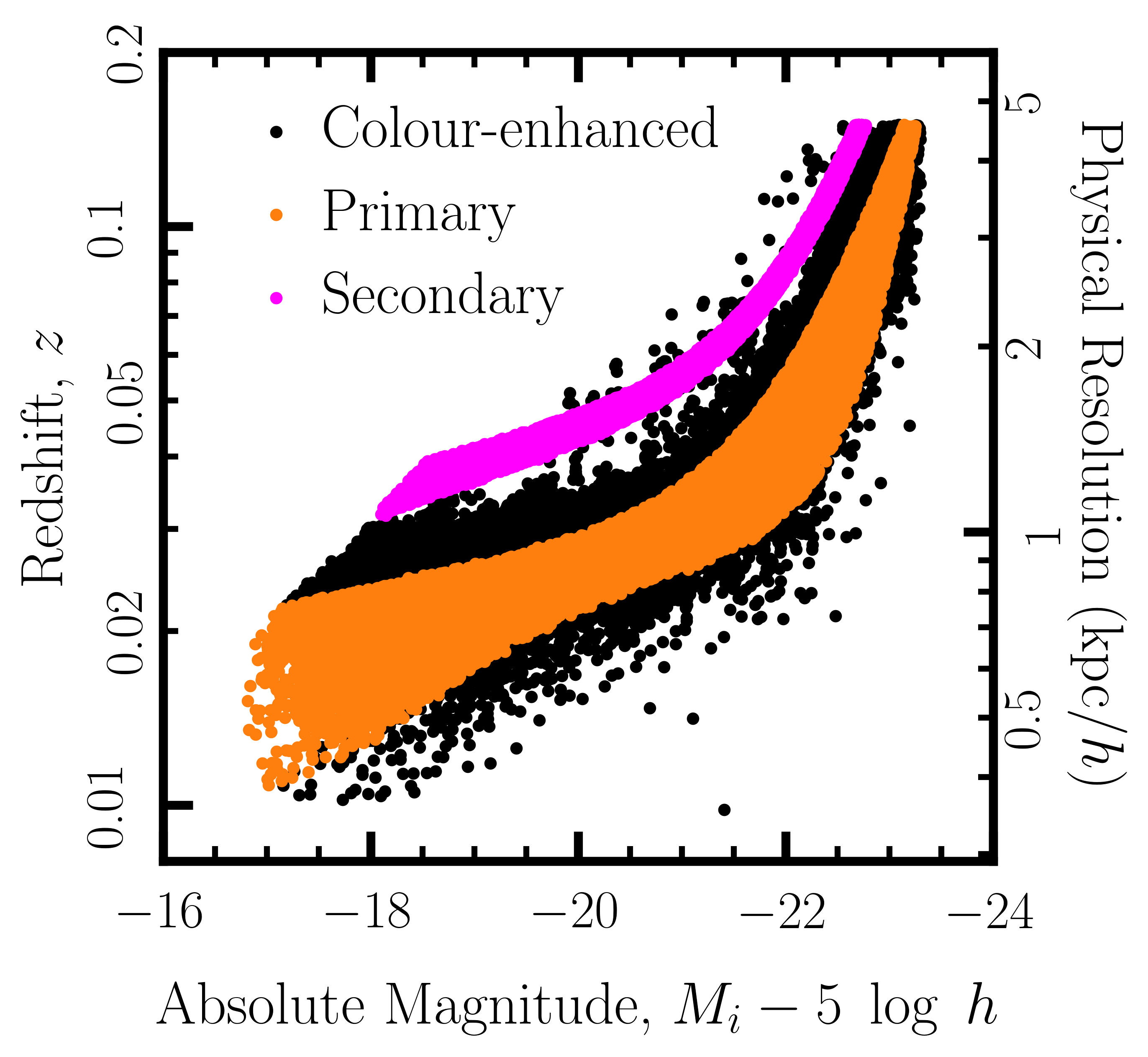}
    \caption[MaNGA selection: redshift-luminosity]{MaNGA survey galaxy selection and samples in the plane of redshift, $z$, and absolute $i-$band magnitude, $M_i$, as described in \cite{2017AJ....154...86W}. Coloured points show individual candidate MaNGA targets from the Primary (orange), Secondary (magenta), and Colour-Enhanced (black) samples of the MaNGA survey. The physical resolution axis assumes the 2.54 arcsecond median effective point-spread function (atmospheric and instrumental) of MaNGA output cubes \citep{2016AJ....152...83L} which is converted to a physical scale at the corresponding redshift.}
    \label{fig:realifs_mangaselection}
\end{figure}

MaNGA galaxies are selected from the SDSS main galaxy sample \citep{2002AJ....124.1810S} following a strategy outlined by \cite{2017AJ....154...86W}. The survey is divided into Primary ($47$ per cent), Secondary ($37$ per cent), and Colour-Enhanced supplement ($16$ per cent) samples. Figure \ref{fig:realifs_mangaselection} shows the MaNGA survey galaxy selection and samples in the plane of redshift and absolute $i-$band magnitude. Galaxies in the Primary sample (orange) are selected such that they are covered out to $1.5$ effective radii, $R_{\mathrm{eff}}$, by the observational footprint of one of the five MaNGA science IFU designs, here denoted: N$19$, N$37$, N$61$, N$91$, N$127$, corresponding to the number of fibres in each bundle. The diameters of the observational footprints of these IFUs are: $12.5,\;17.5,\;22.5,\;27.5,\;\mathrm{and}\;32.5$ arcseconds, respectively, after dithering. The Colour-Enhanced sample (black) is a supplement to the Primary sample that includes galaxies from poorly sampled regions of colour-magnitude plane including low-luminosity red galaxies, high-luminosity blue galaxies, and green valley galaxies. IFU design allocation for the Colour-Enhanced sample is the same as for the Primary sample -- covering galaxies out to $1.5\;R_{\mathrm{eff}}$. The Secondary sample covers galaxies in the same luminosity range as for the Primary sample but at slightly higher redshifts. Galaxies in the Secondary samples are assigned IFU designs such that they receive greater coverage, out to $2.5\;R_{\mathrm{eff}}$.

In our synthetic MaNGA stellar kinematic observations, we marginalize over the luminosity axis of Figure \ref{fig:realifs_mangaselection} and instead select two redshifts at which to mock-observe each TNG50 galaxy sightline. Since our LOSVD cubes are generated in the rest-frame in physical coordinates, the simulation snapshot redshift for our TNG50 sample, $z=0.1$, is irrelevant and each sightline can be mock-observed at any arbitrary redshift. Specifically, for each of the four sightlines for every TNG$50$ galaxy, a redshift is drawn from the marginal redshift distributions of each of the Primary and Secondary samples in Figure \ref{fig:realifs_mangaselection}, for a total of $893\times4\times2=7144$ synthetic MaNGA stellar kinematic observations. These redshifts set the physical-to-angular scale of the LOSVD cubes for subsequent convolution with atmospheric seeing, IFU assignment, and observation. Simultaneously matching to the luminosities of MaNGA galaxies would result in the rejection of TNG50 galaxies from our sample. Instead, synthetic observations are produced for every TNG50 galaxy. In this way, matching of real and synthetic redshift-luminosity distributions of via rejection of either MaNGA or TNG50 galaxies can be carried out afterward, if desired.

\subsubsection{Atmospheric seeing selection}\label{sec:realifs_seeing}

Galaxy light is spatially broadened by the atmospheric seeing before any interaction with ground-based instruments. Consequently, selection and convolution with the atmospheric seeing directly follows redshift selection. Given a redshift for a particular synthetic observation, the physical FOV of the corresponding idealized synthetic LOSVD datacube is converted to an angular size. Then, an atmospheric Point-Spread Function (PSF) is drawn randomly from the distribution of atmospheric seeing full-width at half-maxima (FWHM, arcseconds) estimates for real MaNGA observations using the \texttt{drpall} tables\footnote{The seeing values used here derive from the MaNGA \texttt{drpall} tables which include minimum, maximum, and median guide-star atmospheric seeing estimates for every combined set of exposures. For a given galaxy, we take the median seeing FWHM over all exposures from these tables. Data model for \texttt{drpall} tables: \url{https://data.sdss.org/datamodel/files/MANGA_SPECTRO_REDUX/DRPVER/drpall.html}}. These atmospheric seeing estimates derive from guide-star observations and are quoted at a reference wavelength, $\lambda_0=5400$ \AA{}, and average $\sim1.5$ arcseconds \citep{2016AJ....151....8Y}. The atmospheric seeing FWHM has a weak $(\lambda/\lambda_0)^{-1/5}$ wavelength dependence \citep{1966JOSA...56.1372F,1978JOSA...68..877B}. The seeing also exhibits focus offsets due to changes in shape of the telescope focal plane with wavelength \citep{2006AJ....131.2332G}. 

Since the input data for our synthetic MaNGA survey are stellar LOSVD cubes, the wavelength dependence in the atmospheric seeing and focus offsets are neglected and the telescope focal plane PSF kernel is constructed from the effective values at the reference wavelength, $\lambda_0=5400$ \AA{} from the \texttt{drpall} tables. This focal-plane PSF (pre-fibre) can be reasonably approximated as a combination of two circular Gaussian profiles \citep{2015AJ....150...19L}:
\begin{align}
\label{eq:realifs_convolution}
k(r) = \frac{9}{13}G(r,\sigma_1) + \frac{4}{13}G(r,\sigma_2)
\end{align}
where
\begin{align}
G(r,\sigma_i) = \frac{1}{\sigma_i \sqrt{2 \pi}} \exp \left( \frac{-r^2}{2\sigma_i^2} \right)
\end{align}
and
\begin{align}
\label{eq:realifs_kernels}
\sigma_1 = \frac{\mathrm{FWHM}_{\mathrm{drpall}}}{1.05 \times 2\sqrt{2\ln2}} \quad \mathrm{and} \quad \sigma_2 = 2\sigma_1
\end{align}

Therefore, given the quasi-randomly selected MaNGA redshift, $z$, and seeing, $\mathrm{FWHM}_{\mathrm{drpall}}$, for the synthetic observation, the atmospheric seeing kernel is constructed following Eqs. \ref{eq:realifs_convolution} and convolved with every velocity channel of the idealized stellar LOSVD cube. 

\subsubsection{Assignment of IFU designs}
\label{sec:realifs_ifuassign}

Following the MaNGA IFU assignment process, IFU designs are assigned to each synthetic observation based on: (1) the galaxy's angular effective radius at the target redshift and (2) the target sample (Primary or Secondary)\footnote{We neglect the constraint from finite number of IFUs of each design. See \citet{2017AJ....154...86W} for an exact description of the MaNGA IFU design selection strategy on a per-plate basis.}. For each ``observation'', the IFU design with the smallest difference between its own observational footprint diameter and $2\times N_{R_{\mathrm{eff}}} \times R_{\mathrm{eff}}$ is chosen, where $N_{R_{\mathrm{eff}}}$ is $1.5$ for the Primary and $2.5$ for the Secondary sample. The stellar half-\emph{mass} radius of the galaxy is adopted as a proxy for a light-weighted $R_{\mathrm{eff}}$ for the TNG50 sample. Galaxies for which $2\times N_{R_{\mathrm{eff}}} \times R_{\mathrm{eff}}$ is larger than the N$127$ design footprint are assigned the N$127$ design. Similarly, galaxies for which $2\times N_{R_{\mathrm{eff}}} \times R_{\mathrm{eff}}$ is smaller than the N$19$ design are assigned the N$19$ design footprint.

\subsubsection{Line-spread function}
The instrumental line-spread function (LSF) describes the spectral response of the telescope optics and spectrograph to incoming light. Consequently, accurate characterization of the LSF is required to estimate intrinsic absorption/emission line widths and stellar LOSVDs -- both of which are convolved with the LSF.  \rifs{} can be used to convolve an arbitrary LSF with the non-spatial axis of the data at various steps. However, convolution with an LSF should be restricted to cases where LSF units are relevant: (1) wavelength/frequency or (2) velocity. The LSF should most realistically be applied after the synthetic fibre measurement step (i.e. after spatial integration of the non-broadened input data)\footnote{However, testing showed that the output data are nearly identical when LSF is applied to either: (a) the input data cube; (b) the fibre measurements; or (c) the reconstructed output cubes.}. \cite{2021ApJ...915...35L} show that the LSF for MaNGA observations is well-characterized by a Gaussian -- for which the $1\sigma$ width is denoted $\sigma_{\mathrm{instr}}$. The same has been shown for the SAMI survey \citep{2017ApJ...835..104V}. The LSF $\sigma_{\mathrm{instr}}$ varies spatially, spectrally, and temporally in MaNGA. We marginalize over these variances by adopting a fixed $\sigma_{\mathrm{instr}}=74$ km s$^{-1}$ for our synthetic stellar kinematic data -- an average $\sigma_{\mathrm{instr}}$ estimated over all wavelengths used for stellar template fitting and all MaNGA spectra \citep{2021ApJ...915...35L}.

\subsubsection{Synthetic MaNGA kinematics}\label{sec:realifs_examples}

Given a redshift and designated IFU design, fibre measurements are applied to the seeing-convolved LOSVD cubes following the method outlined in Section \ref{sec:realifs_fibreobserve}. The fibre LOSVDs are then convolved with the LSF and subsequently spatially reconstructed onto an output cube with spatial scale 0.5 arcseconds/spaxel. The spatial reconstruction follows the modified Shepard method using Gaussian weights with $(\sigma,r_{\mathrm{lim}})=(0.7,1.6)$ arcseconds as described in Section \ref{sec:realifs_grid} and used by the MaNGA DRP \citep{2016AJ....152...83L}. This procedure is carried out for all TNG50 galaxy sightlines for both the Primary and Secondary samples yielding a total of $893\times4\times2=7144$ synthetic stellar kinematic cubes. Both the stellar LOSVD cubes and intermediate LSF-convolved fibre LOSVDs are made publicly available.

\begin{figure*}
	\subfloat{\includegraphics[width=0.95\linewidth]{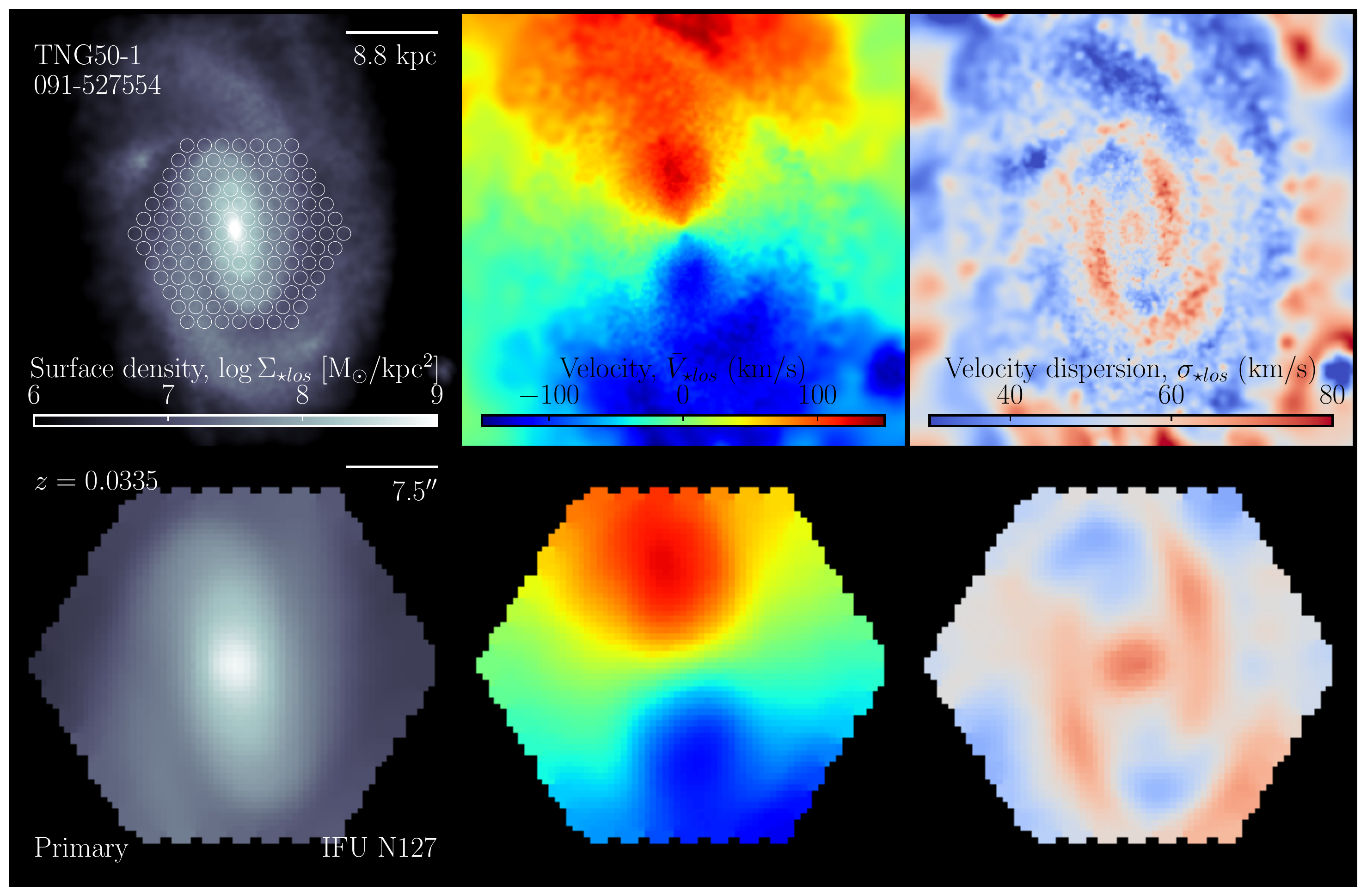}}\\
	\vspace{-16pt}
	\subfloat{\includegraphics[width=0.95\linewidth]{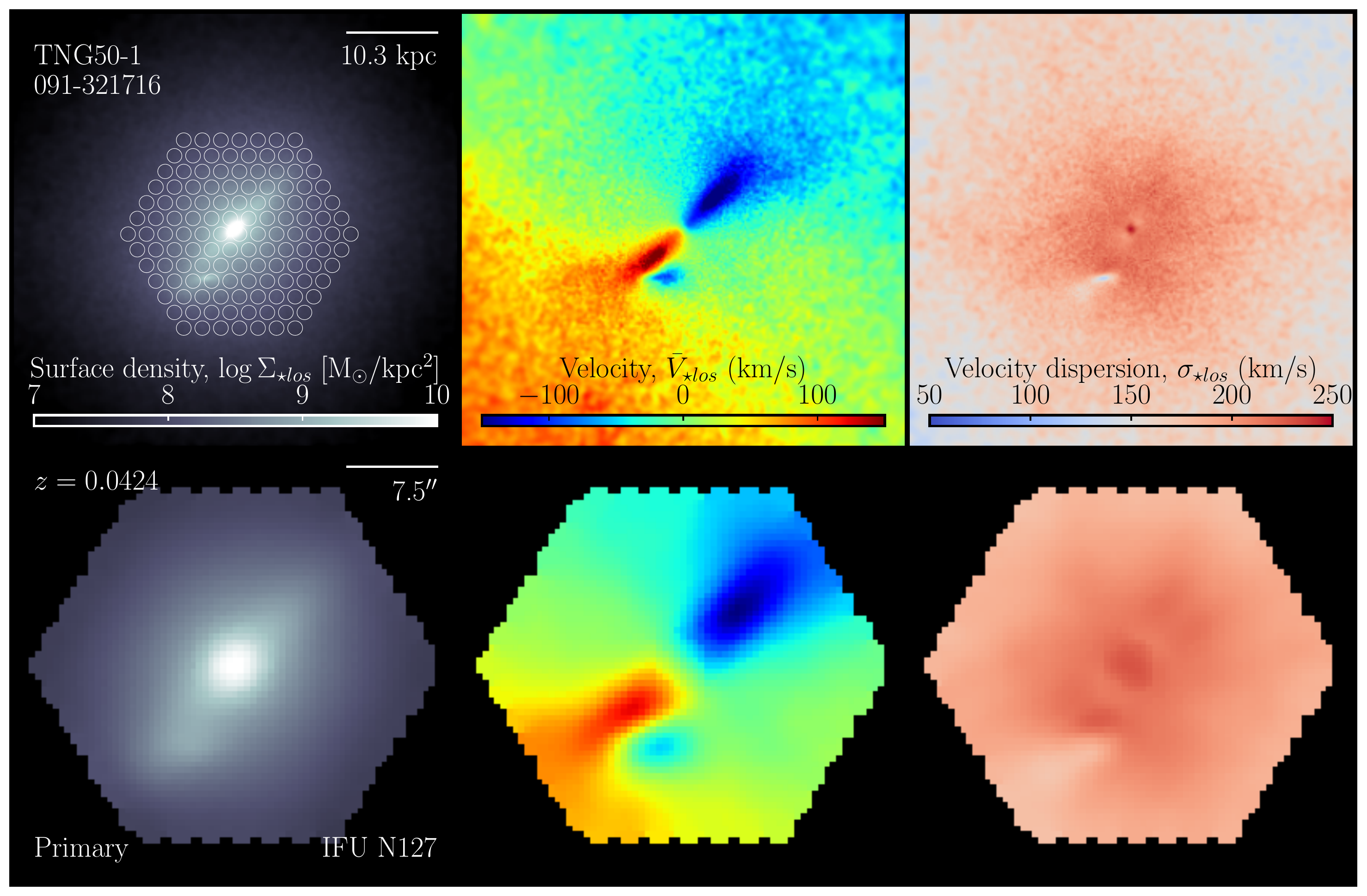}}
   \caption[TNG50 MaNGA Examples]{Construction of MaNGA stellar kinematic cubes for six galaxies from the MaNGA stellar kinematic survey of TNG50 (\emph{continued on next page}). The first four galaxies are mock-observed at redshifts drawn from the MaNGA Primary sample. The last two galaxies are mock-observed at redshifts from the Secondary sample. For each galaxy, the upper row of panels shows the idealized stellar mass surface density, velocity, and velocity dispersion maps as in Figure \ref{fig:realifs_ideal}. The stellar mass surface density map shows the physical footprint for the first of three dithered MaNGA IFU exposures at the selected redshift. The lower row of panels show maps computed from the output cubes after: (1) convolution with the atmospheric seeing; (2) assignment of an IFU design; (3) registration of flux in the IFU fibres; (4) convolution with the line-spread function; and (5) spatial reconstruction onto a Cartesian grid. The redshift, target sample, IFU design, and angular scale are indicated in the synthetic MaNGA stellar mass surface density maps.}
    \label{fig:realifs_exs}
\end{figure*}

\begin{figure*}
\ContinuedFloat 
	\subfloat{\includegraphics[width=0.95\linewidth]{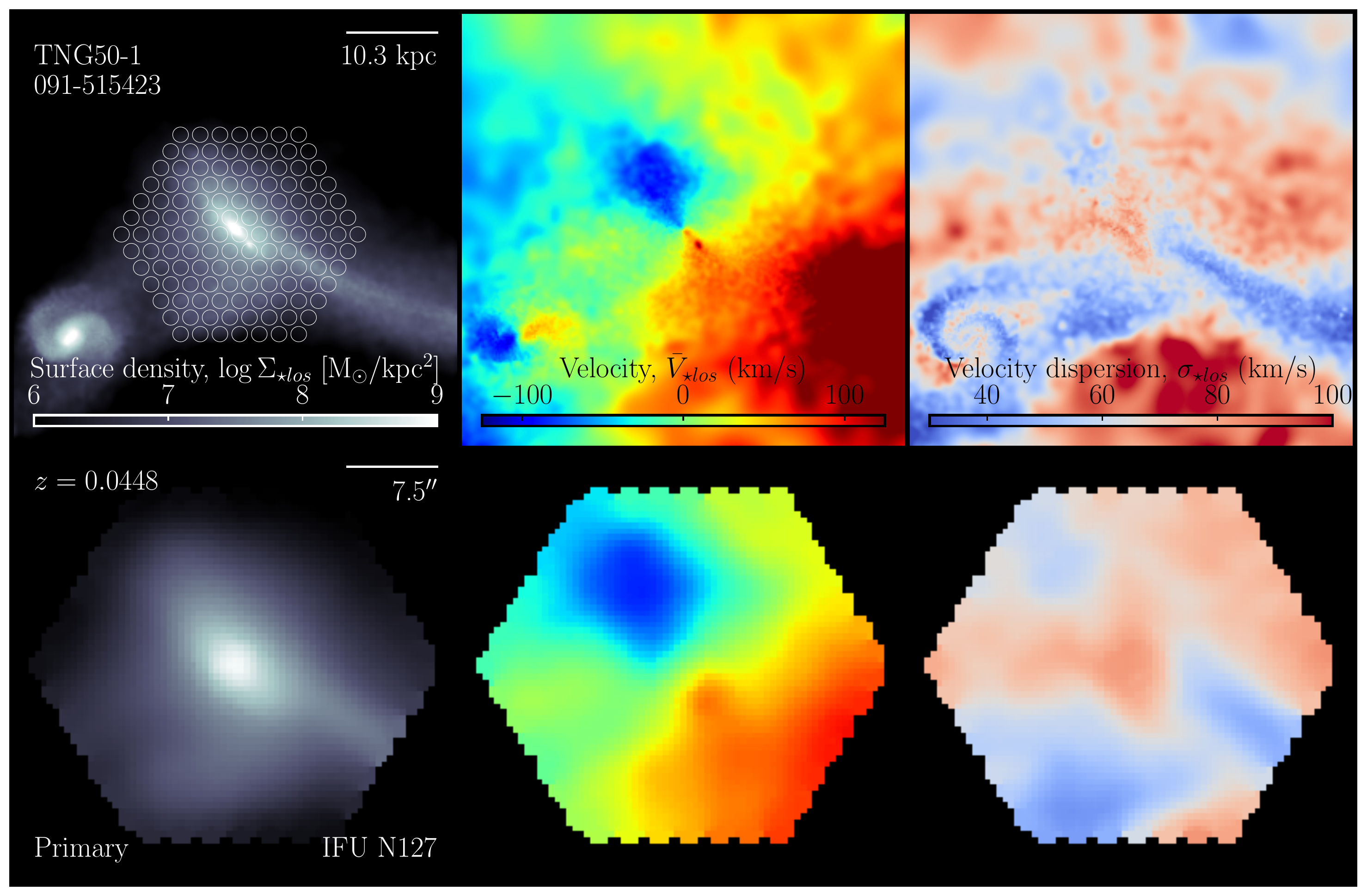}}\\
	\vspace{-16pt}
	\subfloat{\includegraphics[width=0.95\linewidth]{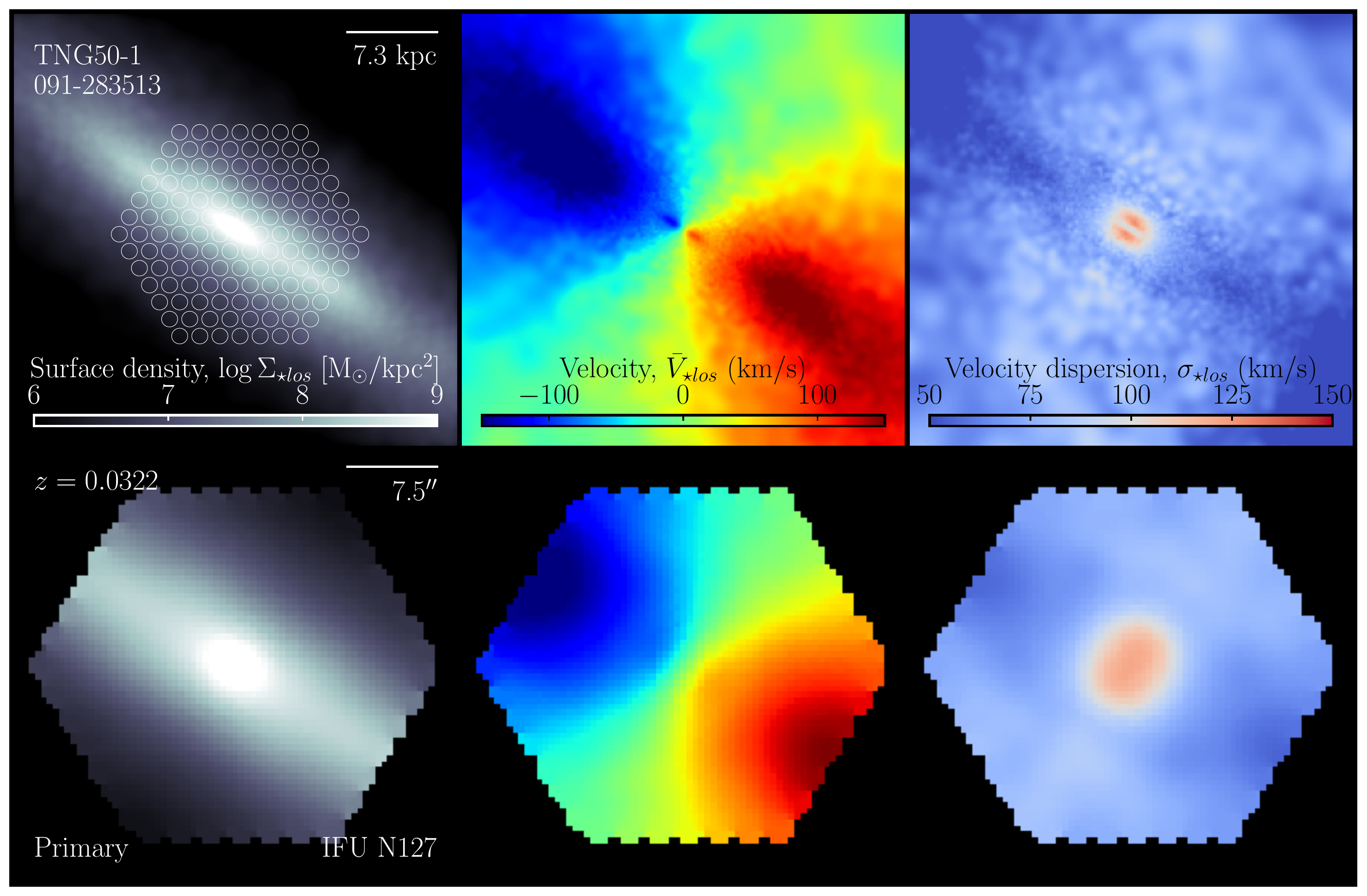}}
   \caption[TNG50 MaNGA Examples]{Construction of MaNGA stellar kinematic cubes for six galaxies from the MaNGA stellar kinematic survey of TNG50 (\emph{continued}).}
\end{figure*}

\begin{figure*}
\ContinuedFloat 
	\subfloat{\includegraphics[width=0.95\linewidth]{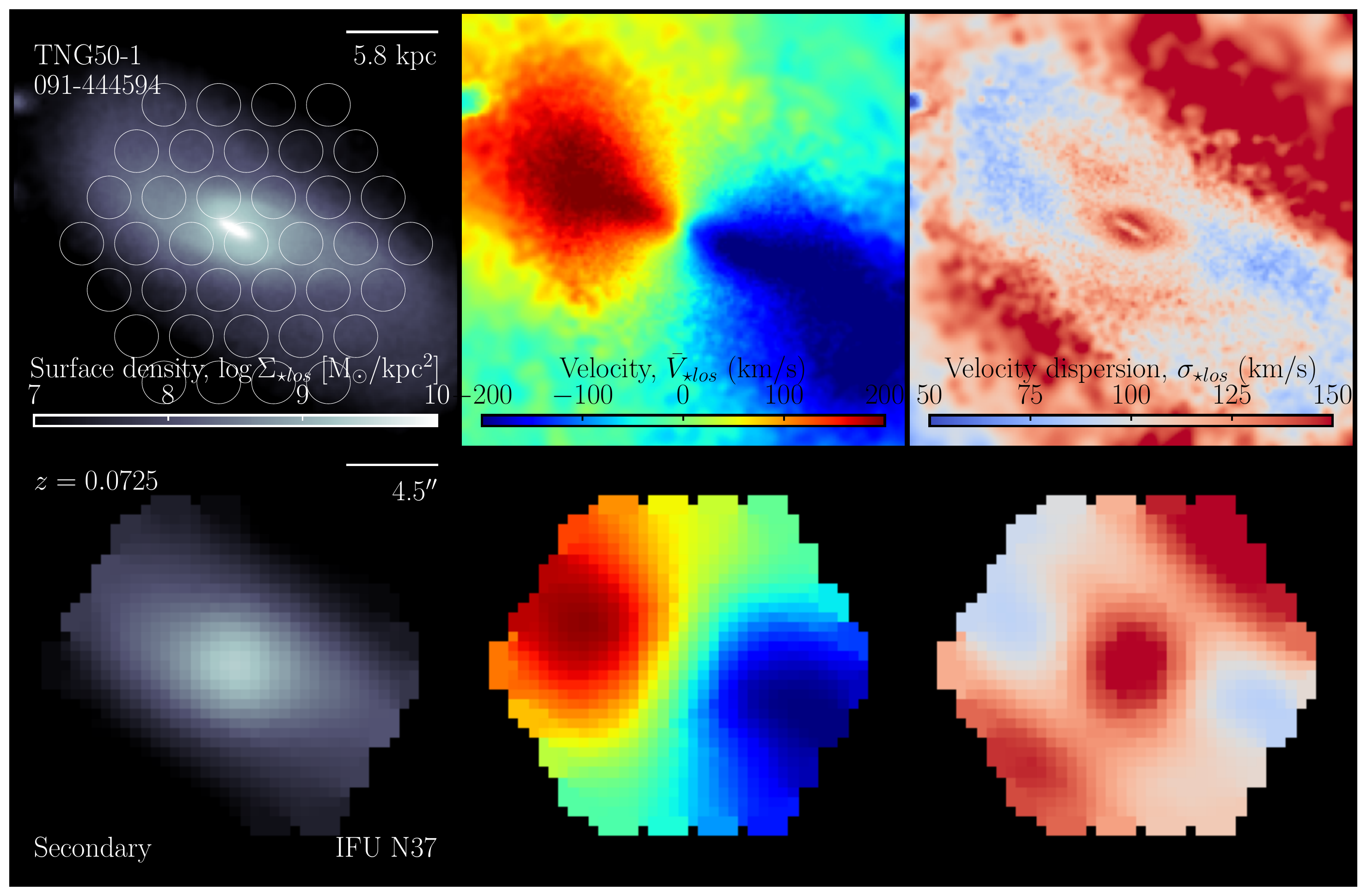}}\\
	\vspace{-16pt}
	\subfloat{\includegraphics[width=0.95\linewidth]{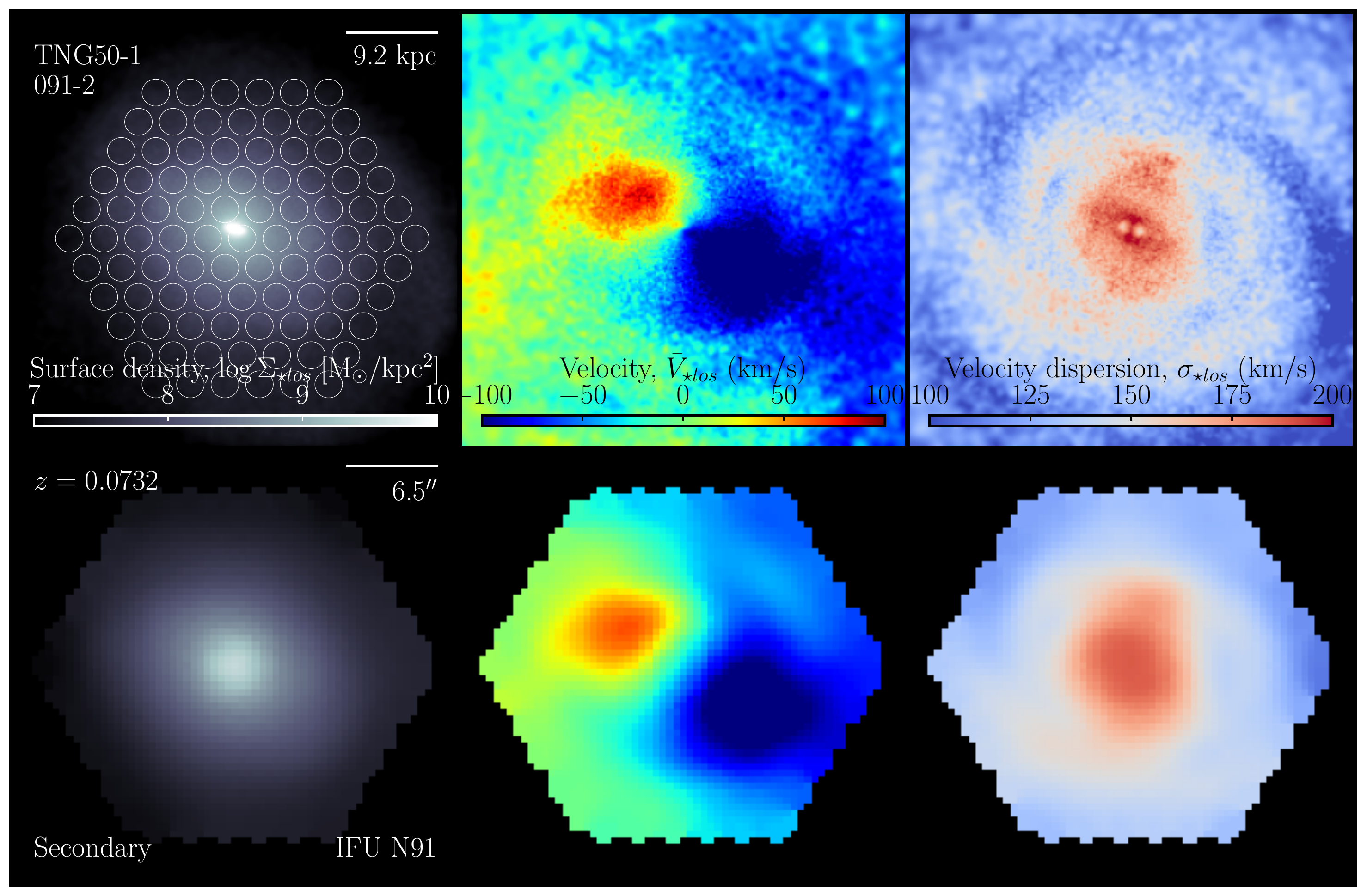}}
   \caption[TNG50 MaNGA Examples]{Construction of MaNGA stellar kinematic cubes for six galaxies from the MaNGA stellar kinematic survey of TNG50 (\emph{continued}).}
\end{figure*}

Figure \ref{fig:realifs_exs} shows six examples of TNG50 galaxies processed with \rifs{} to produce synthetic MaNGA stellar LOSVD cubes. The upper rows for each target show moments derived directly from the idealized input cubes similar to Figure \ref{fig:realifs_ideal} as described in Section \ref{sec:realifs_losvd}. Inset in each stellar mass surface density map is the footprint of the MaNGA IFU design (one of three dithered exposures) assigned to the galaxy based on its selected redshift. These panels show the moments of the LOSVD at their intrinsic spatial and velocity resolution (i.e. before incorporating the atmospheric and instrumental responses to the spatial and velocity dimensions of the data). The lower rows for each target show the same moments computed from the cubes processed with \rifs{} as part of the MaNGA stellar kinematic survey of TNG50. 

The final data products for the MaNGA stellar kinematic survey are as follows in heirarchical data format 5 (HDF5):
\begin{enumerate}[1.]
\item $7144$ two-dimensional matrices of shape $N_{\mathrm{fib}} \times N_v$ containing the LSF-convolved stellar LOSVDs registered by each fibre for a corresponding synthetic observation. $N_{\mathrm{fib}}$ is three times the number of fibres in bundle assigned to each target (three exposure). Each of the 3572 galaxy sightlines is mock-observed at a redshift drawn from both the Primary and Secondary MaNGA target samples. The total data volume of the row-stacked LOSVD data is 2.7 GB and the sizes of files range from 124 kB (N19 design) and 777 kB (N127) and include all details on the observational parameters: redshift, atmospheric seeing, IFU design, LSF width, etc.
\item Stellar LOSVD cubes reconstructed from each of the 7144 row-stacked LOSVD file in (1). The cubes have angular resolution 0.5 arcsecond/spaxel and FOV sizes that are optimized to each IFU footprint. The spatial reconstruction weights for each fibre (Equation \ref{eq:gaussian_weights}) are not included for storage purposes but can be exactly reproduced using the ancillary data included with the cubes (e.g. for propagation of imposed variances on the fibre LOSVDs). File sizes range from to 454 MB (N19 design) to 4.30 GB (N127). Details specific to the reconstruction are appended to the ancillary information in the row-stacked LOSVD files and included with each cube.
\end{enumerate}

\subsubsection{Important considerations and limitations}
\label{sec:limitations}

The \rifs{} software is designed to be sufficiently flexible that it can (1) be adapted to any survey, existing or anticipated, and (2) accept input data cubes which are not constraining in the choice of the non-spatial dimension. The demonstrations above use the idealized stellar velocity cubes. But the non-spatial axis of the input can be any binnable physical characteristic of the relevant particle type. Cubes of stellar age and metallicity, for example, are programmatically identical to velocity cubes with respect to incorporating an IFU's spatial response because each non-spatial channel is handled independently. In contrast, the response of the line-spread function cannot be incorporated into stellar age cubes in a way that is programmatically consistent with velocity or wavelength cubes. Similarly, incorporation of realistic variances in the velocity data used here would differ greatly from how variances would be introduced in wavelength data. Since \rifs{} is designed to process cubes of many types with the response of any IFS instrument, there is no universal approach to incorporating realistic variances.

Consequently, the mass-weighted stellar LOSVD data presented here incorporate the sampling mechanics of MaNGA but do not contain noise. Therefore, algorithms aimed at homogenizing signal-to-noise (S/N) through adaptive spatial binning (e.g. \citealt{2003MNRAS.342..345C}) should focus on homogenizing stellar mass (as signal) -- the units of the cubes and row-stacked LOSVDs. A redshift-dependent noise term can be invoked for more realistic scaling of S/N with redshift. For example, treating angular stellar mass surface density (M$_{\odot}$ arcsec$^{-2}$) as bolometric surface brightness yields the usual $(1+z)^{-4}$ scaling of signal with redshift. Therefore, for the purposes of discarding spaxels with insufficient S/N or for adaptive spatial binning, adopting angular stellar mass surface density as signal (M$_{\odot}$ arcsec$^{-2}$) and noise that increases as $(1+z)^4$ could reasonably account for changes in S/N with redshift. Another approach could be to add noise to the synthetic stellar LOSVD moment maps to match empirical relationships between these moments, their uncertainties, and the target redshift in real survey spaxels.

An observationally realistic treatment of noise in spectrum-derived quantities can only be achieved with light-weighted wavelength cubes produced with dusty, kinematic radiative transfer as input (e.g. \citealt{2021ApJ...912...45N,2022arXiv220311575N}). With such input cubes, the noise characteristics of real survey data can be introduced directly into synthetic fibre spectra. Using the CCD characteristics and wavelength-dependent flux calibration data for real IFS exposures, the calibrated synthetic fibre spectra can be converted to analog (electron) units to obtain source Poisson noise. Sky noise, readout noise, and dark current can be derived from blank-sky exposure spectra -- generally taken concurrently with observations and during commissioning of IFS instruments. By incorporating these components into the synthetic fibre spectra and constructing the noise spectra, both can be spatially reconstructed with \rifs{} to produce survey-realistic cubes. Beyond ensuring that S/N and covariances between spaxels are characterized, the primary advantage of this approach is that the uncertainties in deriving physical characteristics properties from spectra are included in those measurements, as they are from real data.

Computing wavelength cubes for our TNG50 sample with dusty, kinematic radiative transfer has a prohibitive computation expense and is beyond the scope of demonstrating \rifs{} by creating a useful public dataset with detailed limitations. Instead, \rifs{} is designed to be easily integrated into any realism pipeline seeking to emulate the response of IFS instruments for simulated data. Lastly, while \rifs{} contains built-in functions for reproducing the fibre footprints and exposure patterns of the SAMI and MaNGA instruments, these are for convenience only and the base version of \rifs{} is not designed to network with the data products of any specific survey (e.g. obtaining blank-sky spectra and flux calibration data) -- though branches of this kind are certainly encouraged. 

\section{Summary}
\label{sec:realifs_summary}

Forward modelling of galaxies from hydrodynamical simulations to realistic synthetic observations enable (1) even-handed comparison of observations to theoretical predictions and (2) mappings between galaxy observables and their evolutionary histories and origins. IFS enables connections to a much larger set of physical observables than are accessible with imaging alone. To enable such connections, we present \rifs{}: a tool for generating realistic synthetic IFS observations of galaxies from hydrodynamical simulations: covering both spatially discretized spectrographs (e.g. image slicers and lenslet arrays) and fibre-based IFUs. The primary goal of \rifs{} is to unify the advances in IFS surveys and computation galaxy astrophysics by enabling forward-modelling of simulation data into synthetic IFS observables. \rifs{} is specifically designed to accurately emulate the instrumental sampling mechanics and variance propagation of existing or anticipated fibre-based IFS instruments. 

\rifs{} accepts data cubes as input for which two dimensions are spatial and the third dimension can be wavelength or any light- or mass-weighted characteristic of the relevant particles (line-of-sight velocity, stellar age or metallicity, gas metallicity, star-formation rate etc.). The core functionalities of \rifs{} for producing the synthetic IFS observations are:

\begin{enumerate}
\item \textbf{Creation or emulation of arbitrary IFU designs and observing strategies} (Section \ref{sec:realifs_ifus}, Figure \ref{fig:realifs_ifus}). \rifs{} accepts an arbitrary set of fibre positions, dithered exposure offsets or rotations, and corresponding fibre core aperture diameters. Built-in functions are included for setting up SAMI and MaNGA IFU designs and observing strategies. \\

\item \textbf{Integration of spatially discretized data into arbitrary fibre apertures} (Section \ref{sec:realifs_fibreobserve}, Figure \ref{fig:realifs_grid}). \rifs{} effectively uses an inverted \emph{drizzle} algorithm to integrate spatially within fibre apertures. Each non-spatial channel is treated independently. \\

\item \textbf{Spatial reconstruction of irregularly sampled fibre data into regular Cartesian grid spaxels} (Section \ref{sec:realifs_grid}, Figures \ref{fig:realifs_slices} and \ref{fig:realifs_exs}). \rifs{} supports both \emph{drizzle} and flux-conserving Shepard spatial reconstruction methods employed in the DRPs of current IFS surveys. 
\end{enumerate}

Beyond these core functionalities, the tool includes functions necessary for incorporating atmospheric seeing and, for wavelength and velocity cubes, the spectral line-spread function. These additional functions support emulation of the spatial and spectral responses of any IFS instrument type -- not limited to fibre-bundle IFUs. Section \ref{sec:realifs_accuracy} and Figure \ref{fig:realifs_slices} showed that \rifs{} exactly reproduces the flux and variance spatial reconstruction components of the MaNGA DRP using real MaNGA fibre data. In Section \ref{sec:realifs_survey}, we demonstrated an application of \rifs{} to producing a synthetic MaNGA stellar kinematic survey of 893 galaxies with $\logMstar>10$ from the the TNG50 cosmological hydrodynamical simulation (Figure \ref{fig:realifs_exs}). This demonstration shows how to emulate the statistics for redshift (Section \ref{sec:realifs_redshifts}), atmospheric seeing (Section \ref{sec:realifs_seeing}), and IFU design assignment (Section \ref{sec:realifs_ifuassign}) for MaNGA samples. We make our synthetic MaNGA survey of TNG50 galaxies publicly available and discuss ways in which realistic noise may incorporated into synthetic IFS data sets in Section \ref{sec:limitations}. 

The tool is highly flexible with respect the instrumental characteristics of the target IFS survey -- supporting fibres and bundles of varying size and numbers within and between individual exposures. \rifs{} can therefore be used to reproduce the instrumental responses of existing multi-object IFS surveys (e.g. CALIFA, SAMI, and MaNGA), forthcoming surveys (e.g. HECTOR), and imagined IFUs. 

\section*{Acknowledgements}
The authors extend their deep gratitude to Sara L. Ellison, Luc Simard, Julie M. Comerford, and David R. Law for helpful discussions and manuscript comments and Nicholas Scott for providing the SAMI survey IFU fibre positions on behalf of the SAMI Survey team. The authors are grateful to the anonymous referee for providing valued suggestions and comments. CB gratefully acknowledges support from the Natural Sciences and Engineering Research Council of Canada (NSERC) as part of their post-doctoral fellowship program [PDF-546234-2020]. Kavli IPMU is supported by World Premier International Research Center Initiative (WPI), MEXT, Japan. MHH gratefully acknowledges support from the William and Caroline Herschel Postdoctoral Fellowship fund. This research was enabled by computational resources provided by Compute Canada on the Cedar cluster (\url{www.computecanada.ca}). Python packages used in this research include \texttt{Astropy} \url{http://www.astropy.org} \citep{astropy:2013, 2018AJ....156..123A} and Numpy \url{https://numpy.org} \citep{harris2020array}. 

\section*{Data and Code Availability}
The MaNGA stellar kinematic survey of TNG50 galaxies described in Section \ref{sec:realifs_survey} is currently available publicly here: \url{http://idark.ipmu.jp/hsc405/MaNGA/}. Additionally, the input data are available here: \url{http://idark.ipmu.jp/hsc405/LOSVD/}. TNG50 simulations (and all other simulations in the TNG suite) are publicly available at \url{https://www.tng-project.org}. The \rifs{} Python code is publicly available here: \url{https://github.com/cbottrell/realsim_ifs}. The \rifs{} Github repository includes a tutorial folder and notebook detailing the construction of a MaNGA stellar LOSVD cube for a TNG50 galaxy. Our code for the creating idealized stellar/gas LOSVD input data is available upon reasonable request to CB or MHH.



\bibliographystyle{mnras}
\bibliography{References} 




\appendix


\bsp	
\label{lastpage}
\end{document}